\documentclass{aa}  
\usepackage{graphicx}
\usepackage{txfonts}
\begin{document} 

   \title{Magnetic fields and hot gas in M\,101
   \thanks{
Based on observations obtained with {\sl XMM-Newton}, 
an ESA science mission with instruments and contributions directly funded by ESA Member States and NASA
    }
    }
\author{M. We\.zgowiec\inst{1}
\and R. Beck\inst{2}
\and M. Hanasz\inst{3}
\and M. Soida\inst{1}
\and M. Ehle\inst{4}
\and R.-J. Dettmar\inst{5}
\and M. Urbanik\inst{1}}
\institute{
Obserwatorium Astronomiczne Uniwersytetu Jagiello\'nskiego, ul. Orla 171, 30-244 Krak\'ow, Poland \\
\email{markmet@oa.uj.edu.pl}
\and Max-Planck-Institut f\"ur Radioastronomie, Auf dem H\"ugel 69, 53121 Bonn, Germany
\and Instytut Astronomii, Wydzia{\l} Fizyki, Astronomii i Informatyki Stosowanej, Uniwersytet Miko{\l}aja Kopernika, ul. Grudzi{\k a}dzka 5/7, 87-100, Toru\'n, Poland
\and European Space Agency, European Space Astronomy Centre (ESA/ESAC), Camino Bajo del Castillo s/n, 28692 Villanueva de la Ca\~nada, Madrid, Spain
\and Ruhr University Bochum, Faculty of Physics and Astronomy, Astronomical Institute, 44780 Bochum, Germany}
\offprints{M. Weżgowiec}
\date{Received; accepted date}

\titlerunning{Magnetic fields and hot gas in M\,101}
\authorrunning{M. We\.zgowiec et al.}

 
  \abstract
   {
   Studies of nearby spiral galaxies in radio and X-ray wavelengths reveal the structure and energy balance of the 
   magnetic fields and the hot interstellar medium (ISM). In some spiral galaxies, large-scale ordered magnetic fields 
   have been found between the spiral stellar arms (the so-called magnetic arms). 
   One of the considered explanations of their origin is magnetic reconnection, 
   which according to theoretical studies can efficiently heat the low-density ISM.
   }
   {
   We present, for the first time, high-resolution C-band (5\,GHz) radio maps of the nearby face-on spiral galaxy M\,101 
   to study the magnetic fields and verify the existence of the magnetic arms. The analysis of the archival XMM-Newton X-ray data 
   is performed to search for signatures of gas heating by magnetic reconnection effects in the disk and the halo of this galaxy.
   }
   {
   We combine the Very Large Array (VLA) and Effelsberg radio maps of M\,101 to restore the large-scale emission 
   lost in the interferometric observations. 
   From the obtained maps, we derive magnetic field strengths and energy densities, and compare them with 
   the properties of the hot gas found with the spectral analysis of the X-ray data.
   }
   {
   Most of the X-ray emission likely comes from the hot gas in the halo of M\,101. Its temperature is highest above the massive 
   stellar arm and an inter-arm region with enhanced polarised radio emission, as well as in the inter-arm area where neither H$\alpha$ 
   nor \ion{H}{i} emission is visible. In regions outside of the spiral arms lower strengths, energy densities and higher orders
   of the magnetic fields were observed.
   }
   {
   Although M\,101 does not possess well-defined magnetic arms, a rudimentary magnetic arm was identified in one of the inter-arm regions. 
   We found weak signatures of additional heating of the ISM there, as well as in the galactic halo, 
   which could be explained by the action of magnetic reconnection.
   }

    \keywords{galaxies: individual: M\,101 -- 
              galaxies: ISM -- 
              galaxies: spiral -- 
              galaxies: magnetic fields
              }

   \maketitle
%

\section{Introduction}
\label{intro}

In the analysis of the properties of the interstellar medium (ISM) in spiral galaxies, the role of the magnetic fields cannot be neglected because they take part in many dynamical processes \citep[e.g.][]{beck04}. Being also a product of star formation, magnetic fields are 
found to be most prominent in galactic spiral arms, where radio emission, both thermal and non-thermal, is brightest.
Nevertheless, in some spiral galaxies, large-scale ordered magnetic fields have been found between the spiral arms \citep[e.g.][]{beck19}.
However, the real nature of these magnetic arms is still a matter of debate.

A possible explanation of magnetic field ordering is magnetic reconnection involving annihilation of anti-parallel magnetic field components.
Magnetic reconnection is a fundamental process in astrophysical plasma \citep[see][for a recent review]{lazarian20}. The classical Sweet-Parker
model relies on the ohmic dissipation of electric currents concentrated in thin current sheets, which form on the interface of two regions with
anti-parallel magnetic fields. The classical model results in slow action of these effects, 
because the small thickness of the naturally formed current sheet becomes a
bottleneck for the out-flowing plasma carrying the reconnected magnetic field lines. A new paradigm \citep{lazarian99} of magnetic reconnection invokes
a turbulent magnetic field, permitting the occurrence of magnetic field dissipation in microscopic current sheets that form due to magnetic field line
wandering. Recent investigations have shown that magnetic reconnection is fast in the presence of turbulence since it does not depend on resistivity
\citep{kowal09,kowal12} and the effective reconnection speed reaches 10\% of the Alfv\'en speed.
We adopt the turbulent reconnection model as the natural regularisation process of turbulent magnetic fields in the ISM. 
In galaxies the turbulent magnetic fields are transported by the gas flow from spiral arms to the inter-arm regions.
It is plausible that magnetic reconnection could occur there, leading to regularisation of the magnetic fields and simultaneously to the conversion of
magnetic energy into thermal energy \citep{hanasz98}.

The energy of an annihilated magnetic field is partially converted to thermal energy in the surrounding plasma. Such heating has been considered in a
number of theoretical studies \citep[e.g.][]{raymond92,tanuma99, reynolds99, tanuma03}. 
The most effective heating is expected for higher strengths of the magnetic field and for the lower densities of the interstellar gas. 
For a magnetic field strength of around 3\,$\mu$G and gas number densities of the order of 10$^{-3}\,{\rm cm}^{-3}$, the gas can be heated 
to 1\,million K \citep{tanuma03}. Therefore, it should be possible to find signatures of the reconnection heating in the X-ray data of spiral
galaxies.
The magnetic reconnection is more efficient in the star-forming regions due to stronger field tangling, but might be difficult to 
detect because high turbulence prevents long-lasting ordering of the magnetic field. Also, the heating is expected to be less 
significant in regions of higher gas density, where the cooling is faster. In the inter-arm regions and in galactic halos, however, 
lower gas densities provide favourable conditions for detecting reconnection heating.

We admit that the transformation of magnetic to thermal energy is a complex process. Following \citet{lazarian99} and \citet{vishniac99}, only a fraction
of magnetic flux is annihilated by ohmic heating of electrons within the reconnection zone. Another part of magnetic energy is converted 
to high-frequency Alfv\'en waves. These waves may play a role in, for example, the anisotropic heating of ions in stellar coronae.
Finally, according to these authors, the rate of energy emission is hard to estimate since the nature of the turbulence in the reconnection 
zone is uncertain. Having
in mind these uncertainties, we do not favour any specific energy conversion mechanism, but rather consider their variety as a plausible source
of heating of the ISM. In the following we shall use the term `reconnection heating' as a set of processes that convert the magnetic energy to thermal
energy.

Nevertheless, detection of an increased temperature or thermal energy of hot gas in areas free from significant star formation is not sufficient 
to associate it directly with magnetic reconnection. As recent theoretical studies show, the ISM expelled from star-forming regions and 
heated via supernova shocks can move in all directions, and to significant distances from its place of origin \citep[up to 10\,kpc;][]{hu19}. 
Another possibility of ISM heating, which is not directly related to star formation and is likely to occur away from star-forming regions, 
is cosmic-ray streaming, as recently presented by \citet{heintz20}. 

Therefore, it seems very important to simultaneously study the properties of the magnetic field that could help to point to 
a more probable heating mechanism. If the temperature increase 
is supposed to be a result of the magnetic reconnection effects, lower energy densities and higher degrees of order of the magnetic fields 
should be observed simultaneously, being the signature of the conversion of the energy of the turbulent magnetic field. 

We already analysed radio and X-ray data for two spiral galaxies in which magnetic arms were observed. In NGC\,6946 we found a
slight increase in temperature of the hot gas in the inter-arm regions when compared to the spiral arms \citep{wezgowiec16}. This was 
accompanied by lower energies and higher degrees of order of the magnetic fields, which suggested detection of magnetic reconnection heating.
A similar analysis of the deep XMM-Newton X-ray space telescope \citep{jansen01} data for a massive starburst galaxy, M\,83, 
suggested that effects of reconnection heating could have been detected also in the galactic halo \citep{wezgowiec20}.

The spiral galaxy M\,101 is a nearby galaxy seen nearly face-on (see Table~\ref{astrdat}). Its disk is strongly lopsided; the eastern 
arm likely bears signs of past encounters with one or more companion galaxies, and together with M\,101 they form 
a galaxy group \citep{mihos13}. The galaxy has been observed at many wavelengths but, likely due to the large angular size, 
no high-resolution data at high radio frequencies have been published. The available low-resolution radio data suggest that 
most of the ordered magnetic fields are found along the spiral arms, while some of the large-scale polarised radio emission was also 
present outside of the star-forming regions \citep{berkhuijsen16}.

In this paper we present for the first time the radio data from M\,101 at 5\,GHz with sub-arc-minute angular resolution.
We also analyse archival XMM-Newton X-ray data that can help to perform studies of possible
gas heating by magnetic reconnection effects, both in the disk and in the halo of M\,101. 

\begin{table}[ht]
        \caption{\label{astrdat}Basic properties of NGC\,5457 (M\,101)}
\centering
                \begin{tabular}{lc}
\hline\hline
Morphological type & SABc       \\
Inclination\tablefootmark{a}       & 30\degr    \\
Diameter D$_{25}$  & 24\arcmin  \\
R.A.$_{2000}$      & 14$^{\rm h}$03$^{\rm m}$13$^{\rm s}$\\
Dec$_{2000}$       & +54\degr 20\arcmin 57\arcsec       \\
Distance\tablefootmark{b} & 7.4\,Mpc    \\
Column density $N_{\rm H}$\tablefootmark{c}& 8.58$\times$10$^{20}$\,cm$^{-2}$\\
\hline
\end{tabular}
\tablefoot{
All data except inclination, distance, and column density are taken from the HYPERLEDA database -- http://leda.univ-lyon1.fr -- see \citet{makarov14}.\\
\tablefoottext{a}{\citet{kamphuis93}.}
\tablefoottext{b}{\citet{kelson96}.}
\tablefoottext{c}{Weighted average value from \citet{hi4pi}.}
}
\end{table}

Using the sensitive radio data in C-band (4.5-6.5\,GHz) and the X-ray emission from M\,101 in this paper, we analyse the properties 
of the hot gas and the magnetic fields in this galaxy. Especially interesting is the search for regions of enhanced polarised radio emission 
between the spiral arms of the galaxy; the so-called `magnetic arms' observed in NGC\,6946 \citep{beck19} and M\,83 \citep{frick16}. 
Such features were not found in M\,101 by \citet{berkhuijsen16} in the sensitive single-dish data, 
probably due to the low resolution of the observations.
The data presented by us should be sufficient to verify this. The comparable spatial resolutions of the radio and the X-ray observations 
presented in this paper will allow us to study in more detail the spiral arms and other regions of M\,101.
The X-ray data will additionally
give clues about the emission from the halo gas and its possible interactions with the magnetic fields.

M\,101 was observed with the Effelsberg telescope at 6.2 and 11.1\,cm by \citet{berkhuijsen16}, who analysed both the total 
and the polarised radio intensity. Although the low angular resolution of the single-dish observations did not allow them to study the galaxy in detail,
thorough analyses performed by the authors revealed the global structure of the magnetic field and provided important information about 
the rotation measures and the depolarisation between the two frequencies. This can be used to obtain information about the magnetic fields along 
the line of sight, that is, the halo component.

To study the X-ray emission from the hot gas in M\,101, high resolution (both spatial and spectral), and high sensitivity to diffuse structures, 
are needed.
In this paper we focus only on the X-ray data obtained by the XMM-Newton X-ray space telescope \citep{jansen01}. M\,101 has been observed 
with this telescope four times. Three observations between 2002 and 2005 were used in a number of papers, the most important of which
(in terms of the analyses presented in this paper), is the work of \citet{warwick07}, who also analysed the diffuse X-ray emission. 
The authors focused on the total X-ray emission and analysed the global spectrum extracted from the central 10\arcmin\,of the galaxy. 
This was likely caused by the relatively low sensitivity of these moderate-length observations, further diminished by the use of the medium filter 
in two of the observations, which reduces the sensitivity for the softest part of energy range accessible to XMM-Newton.
The most recent data (December 2018) are the sensitive thin-filter observations, aiming to study  the ultra-luminous X-ray 
source (ULX) population in the disk of M\,101.
At the moment of the writing of this paper, they remain unpublished. 

We also note that M\,101 was observed multiple times with the CHANDRA X-ray Observatory for a total of 1\,Ms. An analysis of these data was presented 
by \citet{kuntz10}, who, similarly to \citet{warwick07}, studied the total X-ray emission from the galaxy. This included a single spectrum extracted 
from almost entire galactic disk, the spectrum of the central bulge, as well as two giant star-forming regions. Because the results of these studies 
correspond well to our analyses, throughout this paper we refer to this work (KS10 hereafter).

\section{Observations and data reduction}
\label{obsred}

In this paper we present new and archival Karl G. Jansky Very Large Array\footnote{https://science.nrao.edu/facilities/vla} (VLA) 
C-band radio observations of M\,101, as well as archival X-ray observations performed with the XMM-Newton telescope. 
For the overlays, the optical image in B-band from the Digital Sky Survey, as well as the H$\alpha$ image by \citet{hoopes01}, have been used.
In the following section, details of the data reduction are presented. 

\subsection{Radio data}
\label{radiored}

\begin{figure}[ht]
\resizebox{\hsize}{!}{\includegraphics[clip]{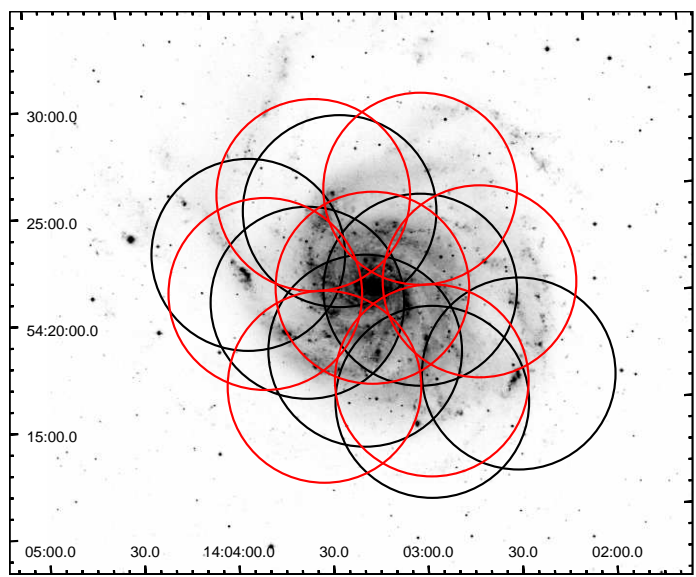}}
\caption{
        Pointings used for the observations of M\,101 presented in this paper. Red and black circles refer to the projects 19B-027 and 15B-292, 
        respectively, and their diameters show the primary beam of the VLA at 5\,GHz.
        }
\label{m101pointings}
\end{figure}

The new C-band D-configuration radio data were observed by the authors within the project 19B-027 in November 2019. Because of the angular size 
of the galaxy, whose radio extent at 5\,GHz is of the order of 15$\arcmin$ \citep{berkhuijsen16}, and the size of the primary beam of the VLA \footnote{https://science.nrao.edu/facilities/vla/docs/manuals/\\oss2019B/performance/resolution} around 
9$\arcmin$ at this frequency, the galaxy had 
to be observed with several offset pointings. We chose a hexagonal layout consisting of seven fields with the inner pointing at the centre of the galaxy. 
The observations were performed in three scheduling blocks (1.5 hours each) consisting of calibrator scans and each of the seven fields covering the galaxy. 
Unfortunately, due to bad weather conditions (heavy snow), one of the observations was not suitable for further processing. As a result, 
the two remaining observations provided 21\,minutes per pointing. To further increase the sensitivity and the coverage of the galaxy we also used archival
C-band D-configuration data for M\,101 obtained within the project 15B-292 that used a different layout of the pointings. 
The observations in this project were scheduled so that each of the fields 
was observed separately for 60\,minutes per pointing. The positioning of all observation pointings used in this paper is presented 
in Fig.~\ref{m101pointings}.

The data were reduced with the Common Astronomy Software Applications (CASA) package\footnote{http://casa.nrao.edu} \citep{mcmullin07}.  
All data affected by shadowed antennas or the radio frequency interference (RFI) were flagged manually. 
In both projects, the radio sources 3C\,286 and J1349+5341 were used for the flux density scale and phase calibration, respectively. 
Because the latter source is unpolarised, it was also used to solve for polarisation leakage. The calibration of the polarisation angle 
was performed with the use of 3C\,286. 

Because the observations of the 19B-027 blocks were not consecutive, we calibrated all three data sets separately. The separate 
calibration of the 15B-292 data sets was straightforward. 
Eventually, we obtained ten calibrated data sets for each of the Stokes parameters. 
To deconvolve the point spread function (PSF) of the interferometer from the dirty map, we used CASA task {\sc tclean}, which  performs 
the Clark CLEAN algorithm \citep{clark80}. For each Stokes parameter, the cleaning runs were performed for all data sets simultaneously 
using the mosaic mode of {\sc tclean}. We find that the best quality images are produced with the use of the robust weighting of the data 
(robust parameter set to one) and the multi-scale CLEAN with scales ranging from 1$\arcsec$ (the cell size) to 600$\arcsec$.
The final images were restored with an elliptical beam of 16$\arcsec\times$13$\arcsec$, which we then convolved to a circular beam 
of 30$\arcsec$ to further improve the signal-to-noise ratio of the faint extended emission.

Despite the high sensitivity of the radio data, due to the large angular extent of M\,101, only little large-scale emission was restored. Therefore, we combined 
our final Stokes I, Q, and U maps with the respective Effelsberg maps at 4.85\,GHz, published by \citet{berkhuijsen16}. 
The combination of data was done using the task {\sc imerg} of the {\sc aips} package\footnote{http://www.aips.nrao.edu}. Because the aim of this task 
was to restore the large-scale diffuse emission that cannot be detected via interferometric observations, we compared the total radio flux densities 
of the combined and the Effelsberg map within the elliptical area encompassing the entire radio emission from M\,101 visible in the latter map.
Table~\ref{vlaeff} presents the comparison of the total flux densities derived from the Effelsberg, the VLA, and the combined maps. 
The total flux densities measured in the Effelsberg and the combined maps agree within 1\%, which confirms the successful merging of the data. At the same time, 
a significantly lower (by almost 60\%) flux density in the VLA maps clearly shows the need to add the single-dish data to interferometric observations 
of spatially extended objects. Nevertheless, no loss of flux was found in the polarised intensity VLA map, which shows the same flux density 
as the Effelsberg map. In the combined map, however, a 10\% higher total flux density was measured. 
Although this still agrees within uncertainties with the total flux densities measured separately in both maps, it is possible that the merging of the maps 
could at least partially restore some of the emission depolarised by the large beam of the Effelsberg telescope
or not detected by the small beam of the VLA.

\begin{table}[ht]
        \caption{\label{vlaeff}C-band total radio flux densities (in mJy) of M\,101.}
\centering
\begin{tabular}{rccccc}
\hline\hline
Map                & Eff\tablefootmark{a}       & VLA   & VLA+Eff       \\
\hline
Total intensity    & 310$\pm$20                 & 127$\pm$6     & 308$\pm$15           \\
Polarised intensity& 28$\pm$3                   & 28$\pm$1      & 31$\pm$2            \\
\hline
\end{tabular}
\tablefoot{
        \tablefoottext{a}{from \citet{berkhuijsen16}}.
}
\end{table}

\subsection{X-ray data}
\label{xrayred}

\begin{table*}[ht]
        \caption{\label{xdat}Characteristics of the XMM-Newton X-ray observations of M\,101}
\centering
\begin{tabular}{rccccc}
\hline\hline
Obs ID          &EPIC-pn filter&EPIC-pn observing mode  &total time [ks] & clean time [ks]\\
\hline
0104260101      & medium       & full frame     &  40.5          &  24.3                  \\
0164560701      & medium       & full frame     &  30.3          &  20.6                   \\
0212480201      & thin         & full frame     &  30.5          &  14.8                  \\
{\it 0824450501}& {\it thin}   & {\it full frame}&  {\it 97.3}   &  {\it 64.9}             \\
\hline
\end{tabular}
\tablefoot{
Only observations marked in italics were used for the spectral analysis (see text for details).
}
\end{table*}

\subsubsection{Data processing and image creation}

As mentioned in Sect.~\ref{intro}, M\,101 has been observed with the XMM-Newton telescope four times. 
For the purpose of this paper, we use all observations to produce images of the diffuse X-ray emission. For the spectral analysis, however, 
only the most recent, sensitive thin-filter data are used. The argumentation for this choice will be presented later in this section. 
Table~\ref{xdat} presents the details of all of the observations. All of the data were processed with the standard procedures of the SAS 19.0.0 package \citep{gabriel04}. 
Calibrated event lists were produced for all three EPIC cameras, using the meta-tasks $emchain$ and $epchain$ for two EPIC-MOS cameras \citep{turner01} 
and the EPIC-pn camera \citep{strueder01}, respectively. To effectively analyse the diffuse emission from the hot gas, all of the data collected during periods 
of intense background radiation needed to be removed. These periods are identified with the use of the light curves of the high-energy emission. 
The light curves allow us to determine the good time interval (GTI) tables that later allow us to filter the event lists for bad data.
For the creation of the GTI tables, the thresholds 0.35\,cts/s and 0.4\,cts/s were used for the light curves (in the 10-12\,keV energy band) of the EPIC-MOS 
and the EPIC-pn cameras, respectively. Such limits are recommended for the analysis of the soft diffuse emission.

Apart from the high-energy background, a significant influence on the faint diffuse emission (especially its spectral analysis) can come 
from the soft proton flare contamination. The filtered event lists were checked for the existence of such contamination 
with the use of a script\footnote{https://www.cosmos.esa.int/web/xmm-newton/epic-scripts\#flare} that applies the calculations developed by \citet{deluca04}.
As a result, medium-filter observations 0104260101 and 0164560701 were found to be slightly contaminated with soft proton flares, 
while the thin-filter observation 0212480201 was contaminated to a very high level. The most sensitive data set, observation 0824450501 
was not contaminated with soft proton flares.

With each of the filtered event lists, we created images and exposure maps (without a vignetting correction) using the 
images script\footnote{https://www.cosmos.esa.int/web/xmm-newton/images}, modified by the authors to allow for adaptive smoothing. 
We used only events with FLAG=0 and PATTERN$\leq$4 (EPIC-pn) or FLAG=0 and PATTERN$\leq$12 (EPIC-MOS), which guaranteed the best quality data 
for the analysis of the diffuse X-ray emission. To detect the entire diffuse X-ray emission, as well as to enable the creation of a hardness ratio map, 
images in two energy bands, 0.2-1 and 1-2 keV, were produced. Because data from four observations were used, we combined the separate images 
into final EPIC images using the SAS task {\sc emosaic}. The combined images were adaptively smoothed with a maximum smoothing scale of 30$\arcsec$, 
and a signal-to-noise ratio of 30. 
This allowed us to reveal the faint diffuse emission and provided a match in resolution with the radio maps. 

\subsubsection{Spectral analysis and background subtraction}
\label{background}

Although still suitable for image creation, observation 0212480201, being very contaminated by soft protons, needed to be excluded from spectral analysis.
To assure the highest sensitivity in the soft energy band, as well as to allow for a direct comparison with our similar studies of other face-on 
spiral galaxies, NGC\,6946 \citep{wezgowiec16} and M\,83 \citep{wezgowiec20}, we used only the EPIC-pn camera for the spectral analysis.
In the direction to M\,101, a relatively high galactic foreground column density $N_{\rm H}$ was measured (see Table~\ref{astrdat}). 
This resulted in a significant absorption that influenced the detection of the softest X-ray emission. This, together with the characteristics 
of the internal background of the pn camera\footnote{https://xmm-tools.cosmos.esa.int/external/xmm\_user\_support\\/documentation/uhb/epicintbkgd.html}, 
led us to only use the data above 0.4\,keV for spectral analysis.

Prior to the spectra extraction, we used the event files and the corresponding calibration data 
of all of the observations to run the {\sc edetect\_stack} 
meta task that performs all stages of the source detection routine, taking advantage of the combined sensitivity of the data.
The meta task was run with the use of standard parameters. The obtained list of detected sources was then used to exclude the detected point-like sources 
from the spectral analysis. The radii of these areas were adjusted to include the brightest emission of the point sources. Because of the extended wings 
of the PSF of the EPIC-pn camera, some residual emission from point sources could still be present in the spectra. 
This, however, was later taken into account by adding to the spectral model a power-law component, which also was used to model the emission 
from potentially unresolved sources. 

Because of the large angular extent of M\,101 (regions used for spectral analysis spread over the central 20$\arcmin$ of the field of view), a proper background 
subtraction becomes challenging. Clearly the use of local background is almost impossible, because it would need to be extracted from areas that 
are significantly off-axis and the potential scaling would introduce unwanted biases.
The often used blank-sky event lists \citep[see][]{carter07}, most likely due to a relatively high column density, led to an overestimated background 
level and the subtracted spectra could not be fit with theoretical models. Having checked that, we decided to account only for the instrumental 
background. We argue that when the external background in the softest energy band is weakened by the higher column density, 
the contribution from the detector might become the most important component of the X-ray background, especially when extended areas 
of low surface-brightness emission are analysed, just like the regions used in our studies. Furthermore, the relatively low sensitivity 
of the analysed data will likely mean that the derived uncertainties of the fitted model parameters will be larger than the influence 
of the external X-ray background.

The instrumental background was modelled with the use of the filter wheel closed (FWC) data\footnote{https://www.cosmos.esa.int/web/xmm-newton/filter-closed}. 
From the repository we chose only those FWC observations that best matched the time of each observation. This was done to best represent the detector 
background, which is known to change over time. The FWC event lists were then filtered in the same way as the source event lists.
Response matrices and effective area files were created for each spectrum with the use of a detector map (image in detector coordinates) that is needed 
for spectral analysis of extended emission. The final spectra, background spectra, response matrices, and effective area files were correspondingly 
merged for all observations with the use of the SAS task {\sc epicspeccombine}. The advantage of this approach, when compared to simultaneous fitting, 
is that all corresponding files are combined first and the background subtraction is performed in the last step, which is especially 
useful in the case of low sensitivity of separate spectra. 

Nevertheless, the final background-subtracted spectra, especially that of the lowest sensitivity, still showed a noticeable contribution from the detector
Al-K$\alpha$ line at 1.5\,keV, as well as an increased level of the softest emission, which was a sign that the background was not entirely removed. 
To look for a solution, we compared the results with spectra obtained from the most sensitive observation only (thin-filter ObsID 0824450501). 
We found that in both cases, the same model fit parameters were obtained, and they were only slightly better constrained in the models 
fitted to the combined spectra. Most importantly, however, the inspection of the thin-filter spectra showed no significant contribution 
from the detector background mentioned before. Therefore, for the spectral analysis presented in this paper, we eventually decided to use only 
the most sensitive thin-filter observation (ObsID 0824450501). 

To further improve the signal-to-noise ratio, spectra were binned. A number of 25 counts per energy bin seems to be a good value for a wide range 
of sensitivities, and was used by us before \citep{wezgowiec16,wezgowiec20}. The spectra were fitted using {\sc xspec12}~\citep{arnaud96}.

\section{Results}
\label{results}

\subsection{Radio emission}
\label{radio}

\begin{figure}[ht]
\resizebox{\hsize}{!}{\includegraphics[clip]{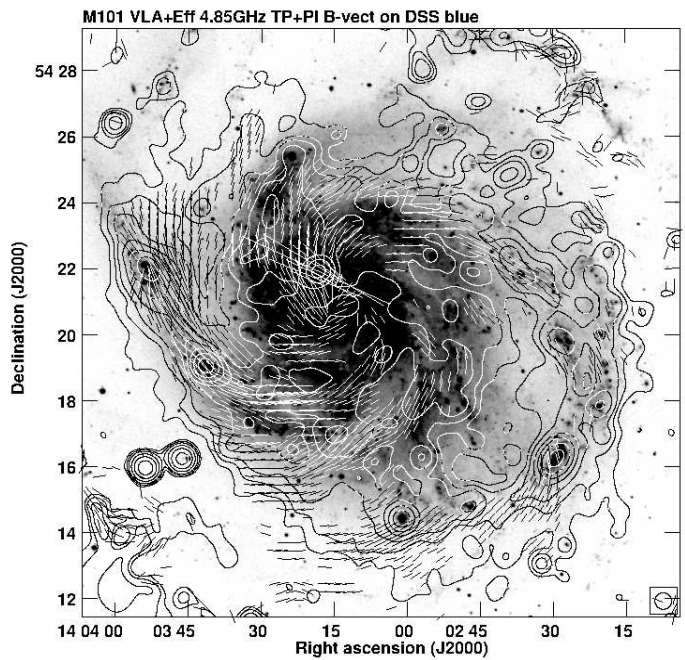}}
\caption{
        Map of total radio intensity at 4.85\,GHz ($\lambda$ 6\,cm) of M\,101 overlaid on the DSS blue map. 
        The contours are 3, 5, 8, 16, 32, 64, 128, and 256 $\times$ 35\,$\mu$Jy/beam. The vectors show the orientation of the magnetic fields 
        and their length of 1$\arcmin$ is proportional to the polarised intensity of 0.17\,mJy/beam. The beam size of 30$\arcsec$ is shown
        in the bottom right corner of the map.
        }
\label{6tp}
\end{figure}

\begin{figure}[ht]
\resizebox{\hsize}{!}{\includegraphics[clip]{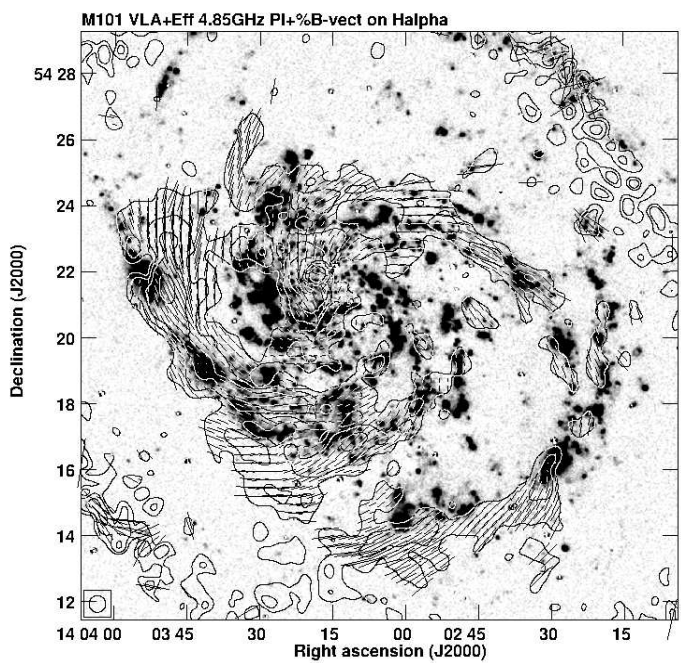}}
\caption{
        Map of polarised radio intensity at 4.85\,GHz ($\lambda$ 6\,cm) of M\,101 overlaid on the H$\alpha$ map. 
        The contours are 3, 5, 8, 16, 32, 64, 128, and 256 $\times$ 15\,$\mu$Jy/beam. The vectors show the orientation of the magnetic fields
        and their length of 1$\arcmin$ is proportional to the degree of polarisation of 50\%. The beam size of 30$\arcsec$ is shown
        in the bottom left corner of the map.
        }
\label{6pi}
\end{figure}

The total radio emission closely follows the star-forming disk of M\,101 with distinct enhancements visible in the spiral arms (Fig.~\ref{6tp}), 
especially the bright H$\alpha$ complexes in the south and at the end of the eastern spiral arm. Northeast from the centre 
(around position R.A. 14$^{\rm h}$03$^{\rm m}$15$^{\rm s}$, Dec +54$\degr$\,22$\arcmin$),
a bright polarised radio source is visible, although its degree of polarisation remains rather low (Fig.~\ref{6pi}). 
The polarised emission also closely follows the star-forming arms with only slight extensions in the south and south-west, as well as the inter-arm 
region between the centre and the eastern arm (around position R.A. 14$^{\rm h}$03$^{\rm m}$45$^{\rm s}$, Dec +54$\degr$\,23$\arcmin$). 
The pitch angle of the magnetic field is relatively low along the spiral arms, but shows a noticeable increase in the southern extensions of 
the polarised emission and in the eastern inter-arm region. 
The polarisation map also shows  a significant asymmetry, with most of the (highly polarised) emission coming from the eastern spiral arm (Fig.~\ref{6pi}). 

Between the centre and the western spiral arm (around position R.A. 14$^{\rm h}$02$^{\rm m}$40$^{\rm s}$, Dec +54$\degr$\,18$\arcmin$), 
an interesting area that does not show any H$\alpha$ or polarised radio emission can be found (Fig.~\ref{6pi}).
A similar feature was also visible in the low-resolution single-dish maps of \citet{berkhuijsen16} and the \ion{H}{i} map, 
in which a significant depression was observed in this area \citep{braun95}.

\subsection{Distribution of the X-ray emission}
\label{dist}

\begin{figure*}[ht]
                        \resizebox{0.5\hsize}{!}{\includegraphics{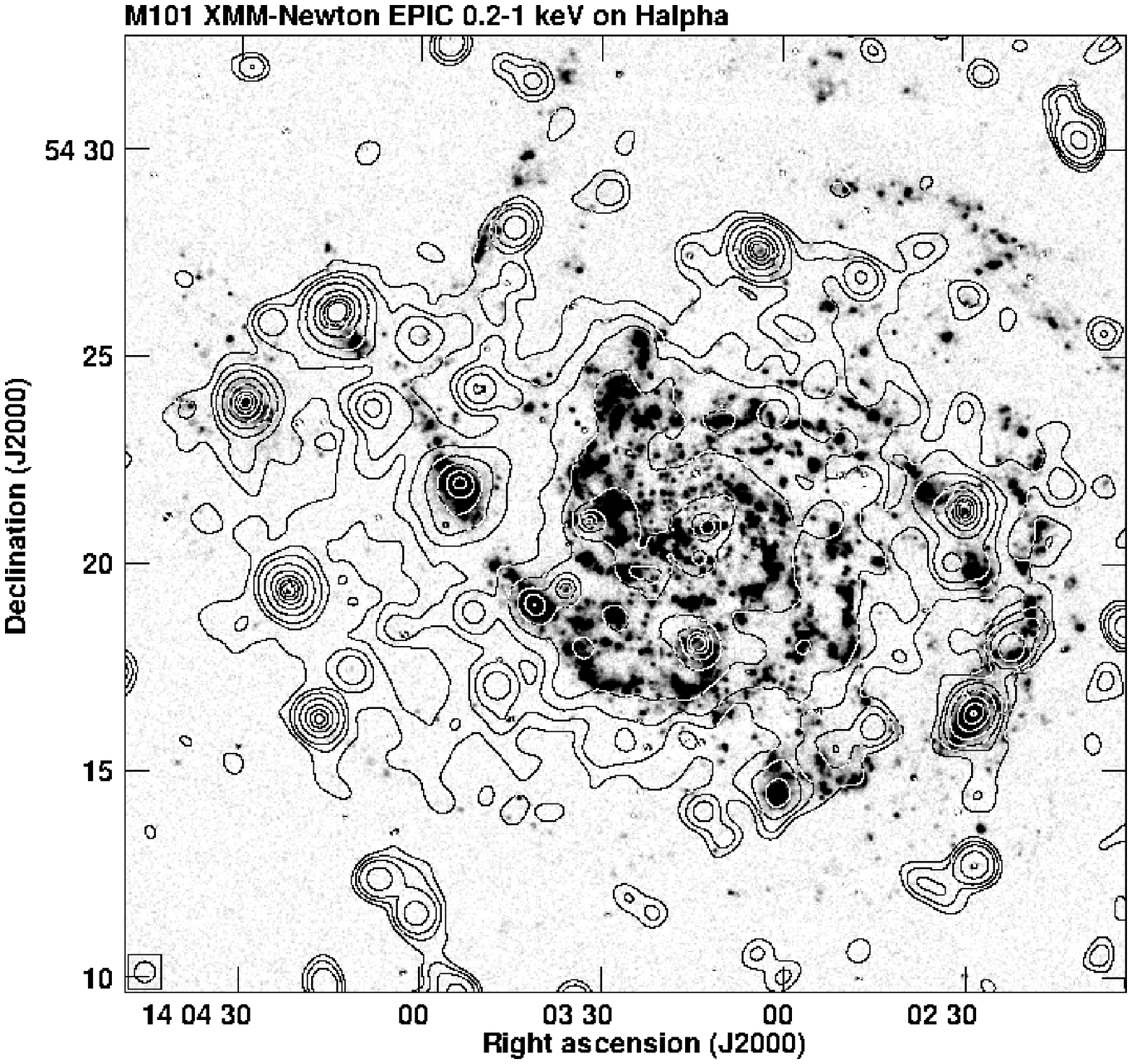}}
                        \resizebox{0.5\hsize}{!}{\includegraphics{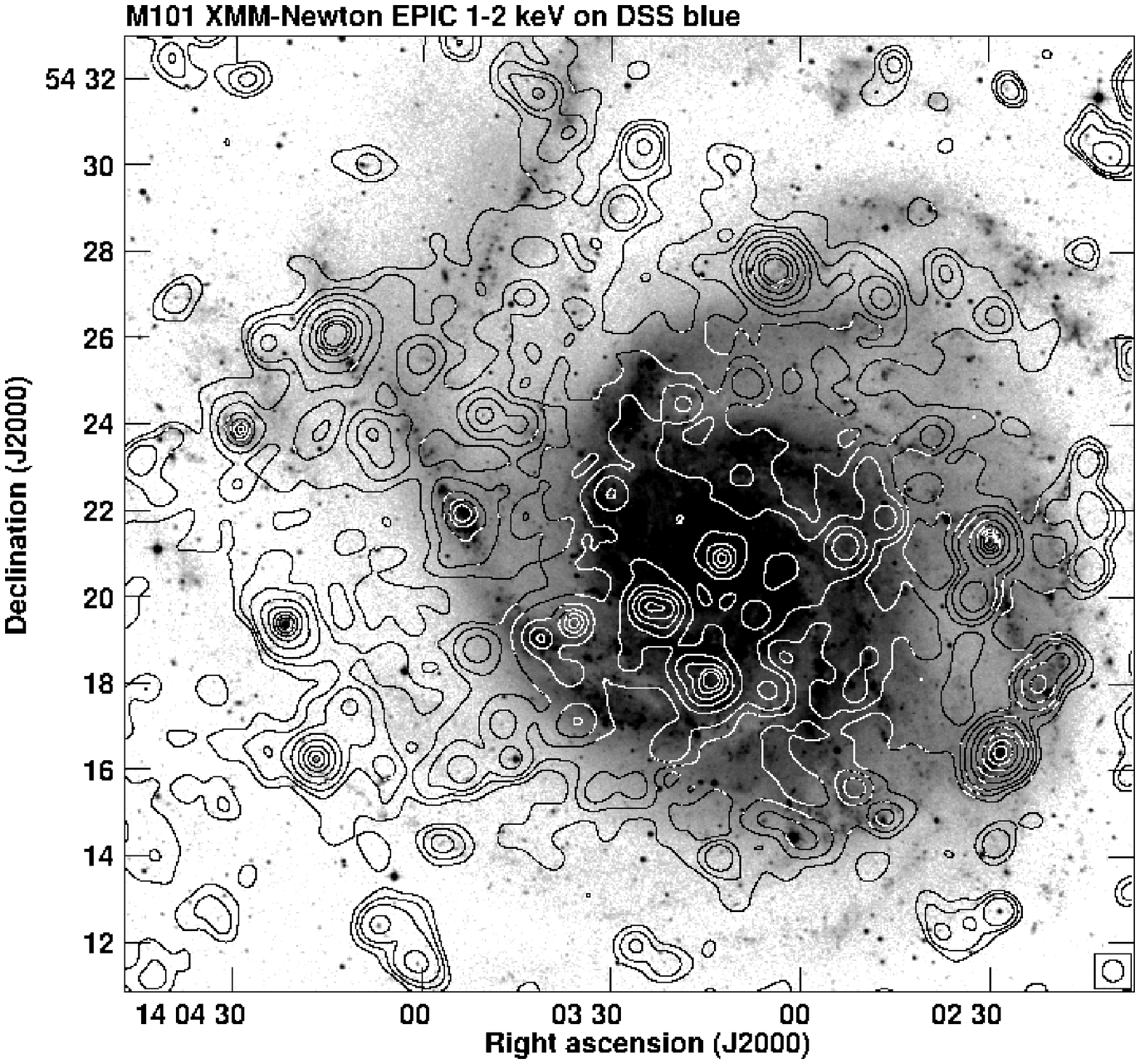}}
                \caption{
		Maps of X-ray emission from M\,101.
		{\it Left:} Soft-band (0.2 - 1 keV) overlaid on an H$\alpha$ image. The
                contours are 3, 6, 12, 24, 48, and 96 $\times$ r.m.s. The map is adaptively smoothed with the
		largest scale of 30$\arcsec$. {\it Right:} Medium-band (1 - 2 keV) overlaid 
                on a DSS blue image.
                The contours are 3, 6, 12, 24, 48, and 96 $\times$ r.m.s. The map is adaptively smoothed with the largest scale of 30$\arcsec$, which
                is shown in the bottom corners of the maps.
                }
                \label{xmaps}
        \end{figure*}

The soft X-ray map (left panel of Fig.~\ref{xmaps}) shows diffuse emission from the entire disk of M\,101.
Contrary to the radio maps, the emission extends significantly beyond 
the eastern spiral arm. Although a number of point sources is visible in this area, this diffuse emission cannot be attributed to the beam-smearing 
effect; the largest scale used in the adaptive smoothing is 30$\arcsec$ (see the beam size symbol in the bottom left corner of the map), 
much smaller than the scale of the emission. The emission of the central star-forming disk (visible in the H$\alpha$ image) is at the level of 24$\sigma$.
The right panel of Fig.~\ref{xmaps} shows that the emission in the higher energy range can be visible to a similar extent than in the 
lower energy range, or even larger (left panel of Fig.~\ref{xmaps}), and the emission reaches 12$\sigma$ at the brightest central part of the galaxy.
This harder emission also seems to be more extended in the area of the eastern inter-arm region, 
where the polarised radio emission can be seen (Fig.~\ref{6pi}).

\begin{figure}[ht]
                        \resizebox{\hsize}{!}{\includegraphics[clip]{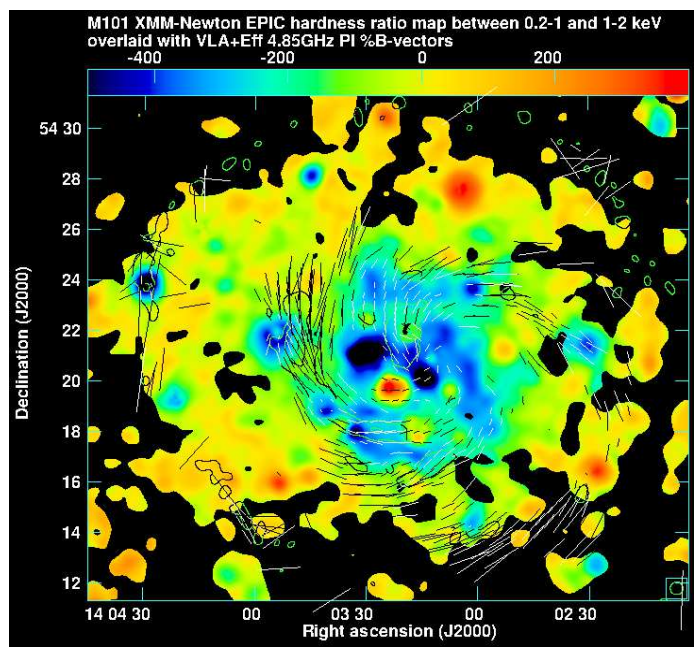}}
                \caption{Map of the hardness ratio between medium and soft bands of the X-ray emission from M\,101 (cf.~Fig.~\ref{xmaps}) 
                overlaid with the B-field vectors proportional to the degree of polarisation. 
                The map is truncated at the 3$\sigma$ level of the medium-band X-ray map. 
                }
                \label{m101hr}
        \end{figure}

To identify areas with significantly different amounts of cooler and hotter X-ray gas, a hardness ratio (HR) map becomes very useful. 
Such a map, presented in Fig.~\ref{m101hr}, is derived from the two maps from Fig.~\ref{xmaps}, as the ratio of their difference and sum. 
The map is presented with the vectors of the magnetic fields proportional to the degree of polarisation. The overall softest emission (shown in blue) 
comes from the most central part of the disk, where the majority of the H$\alpha$ emission can be seen.
Although no clear correlation between the hardness of the X-ray emission and the polarised radio emission can be seen, the areas of the highest degree 
of polarisation tend to be located just outside of the softest emission. Still, most of the diffuse X-ray emission shows the hardness ratios not higher 
than 0.1, which identifies the emission as coming mainly from the soft component.

\subsection{Spectral analysis of the X-ray emission}
\label{spectra}

To investigate the properties of the hot gas in regions of significant polarisation of radio emission, we used the polarisation intensity map 
together with the H$\alpha$ map (Fig.~\ref{6pi}) to select separate regions for the spectral analysis. 
This approach is similar to what was presented by us for other spiral galaxies: NGC\,6946 \citep{wezgowiec16} and M\,83 \citep{wezgowiec20}.  
However, in M\,101 some of these regions correspond to significant H$\alpha$ emission. This is discussed in Sect.~\ref{hotfields}. 

\begin{figure*}[ht]
\resizebox{0.495\hsize}{!}{\includegraphics[clip]{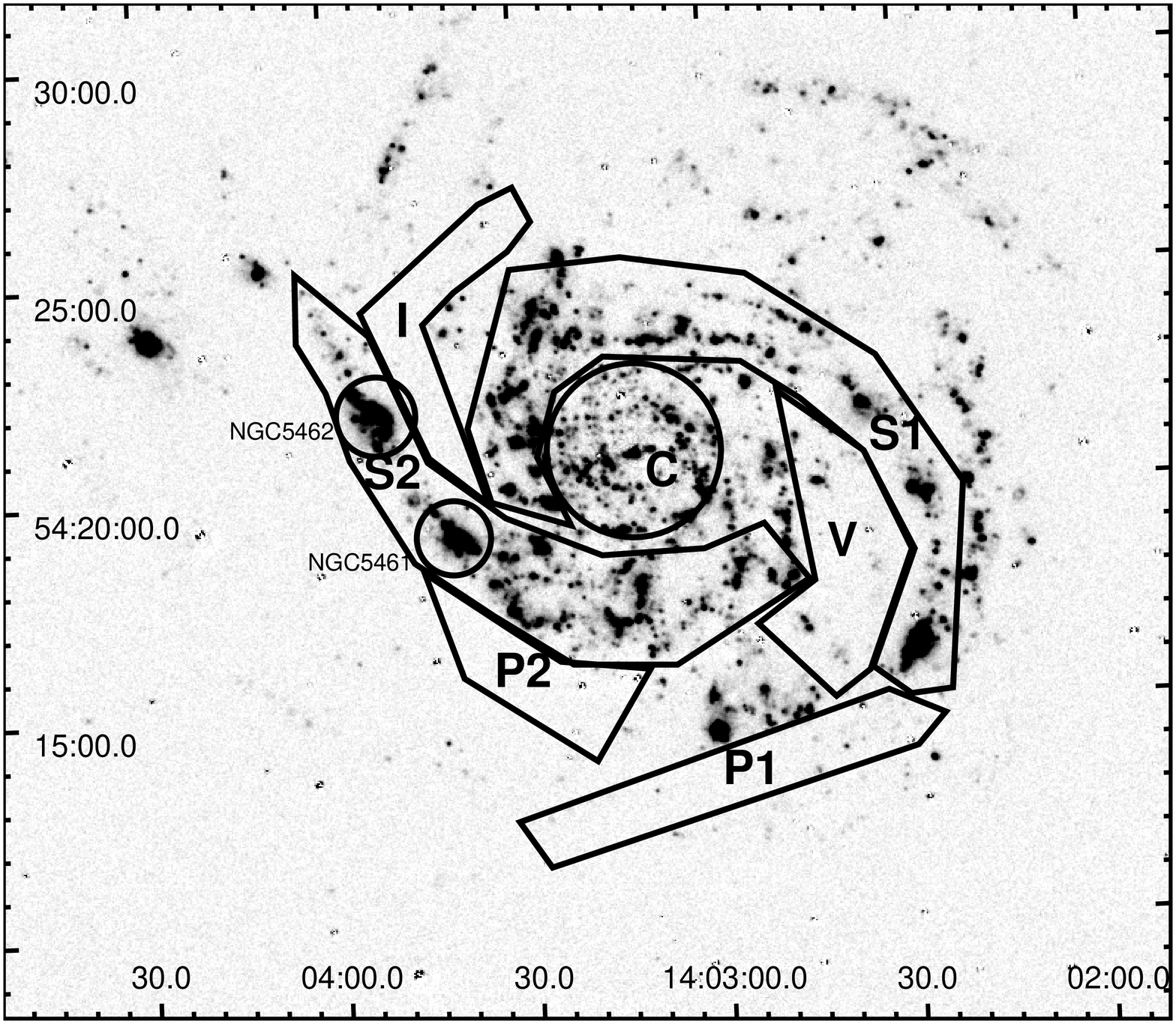}}
\resizebox{0.505\hsize}{!}{\includegraphics[clip]{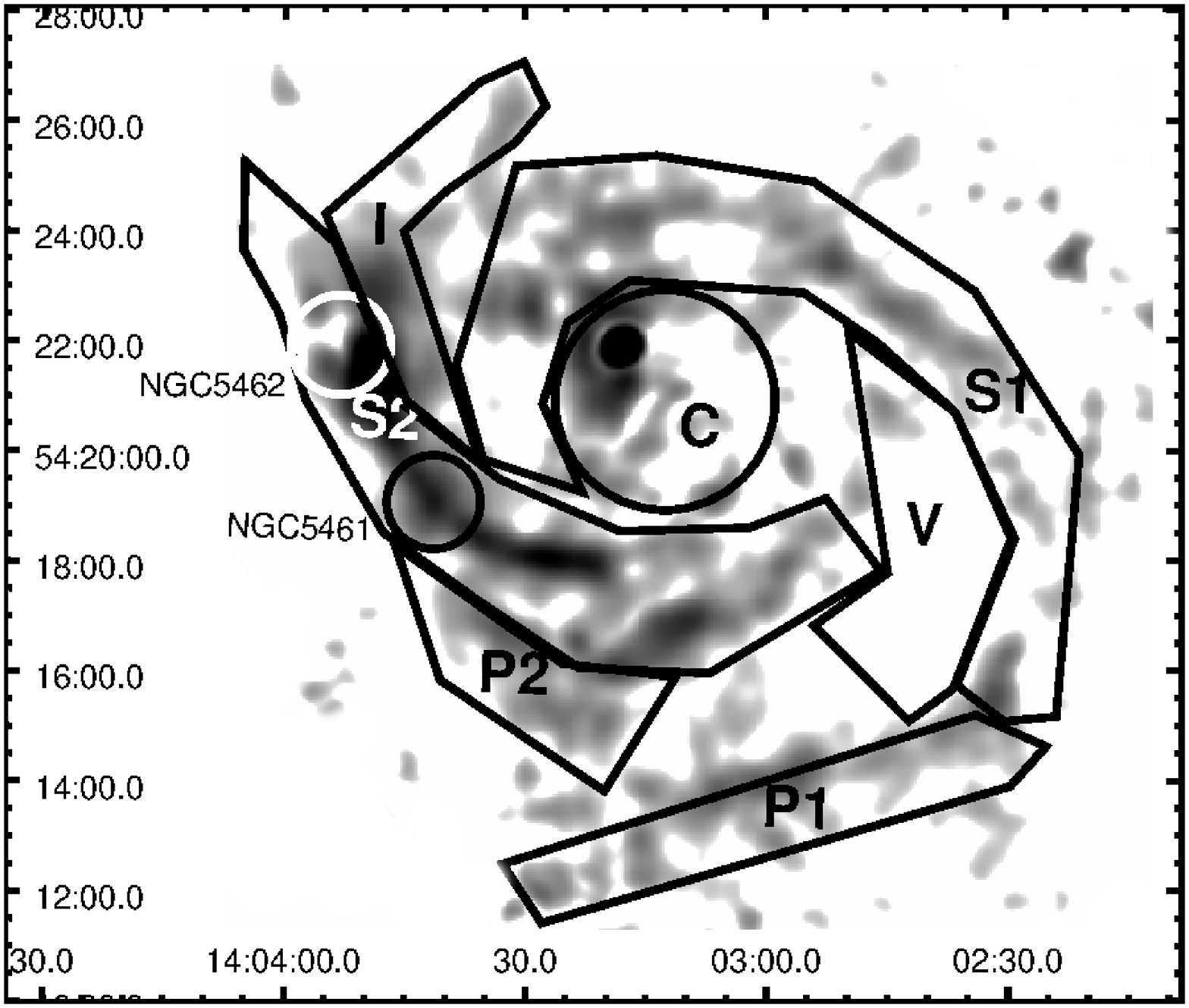}}
\resizebox{0.565\hsize}{!}{\includegraphics[clip]{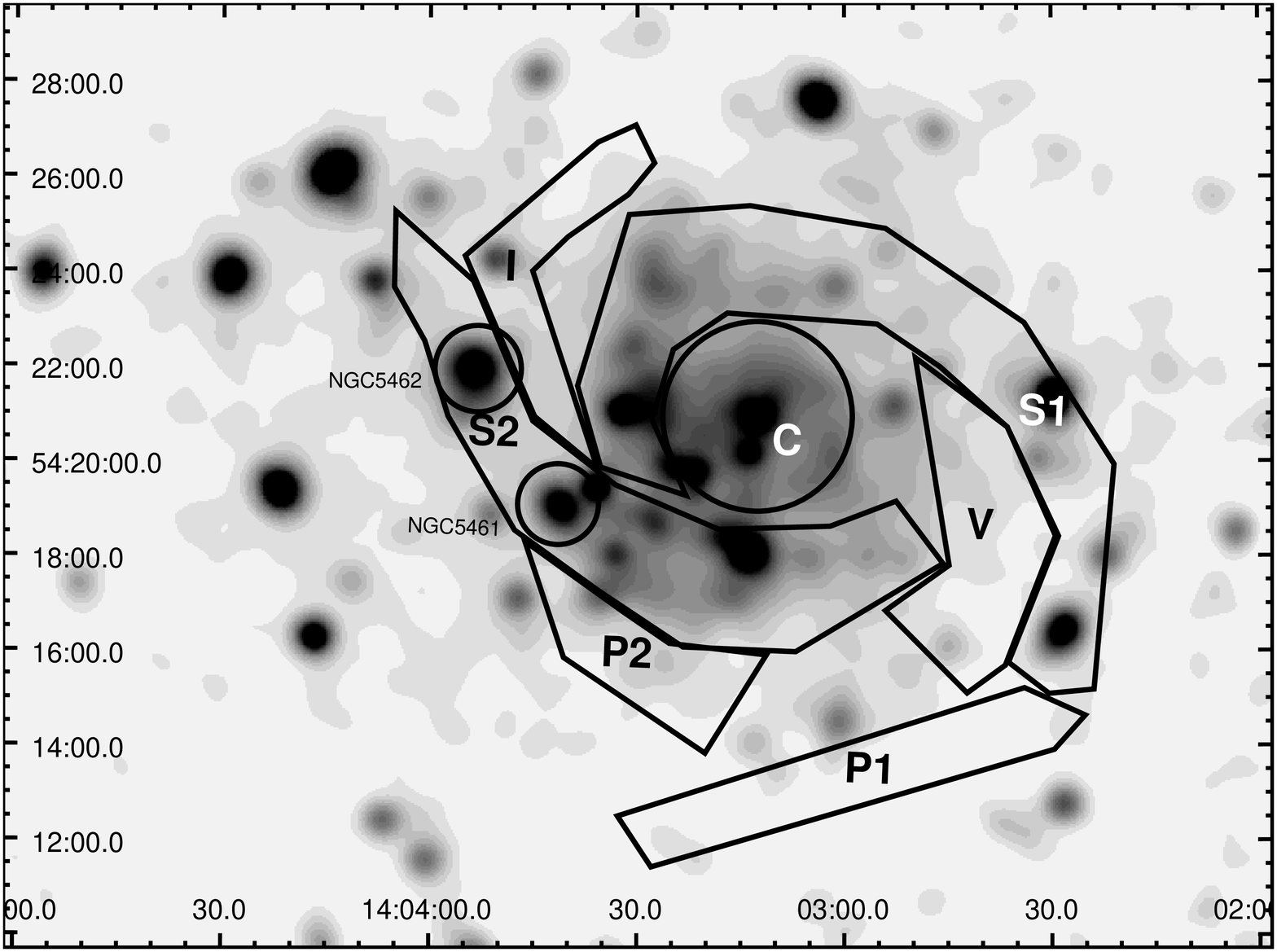}}
\resizebox{0.435\hsize}{!}{\includegraphics[clip]{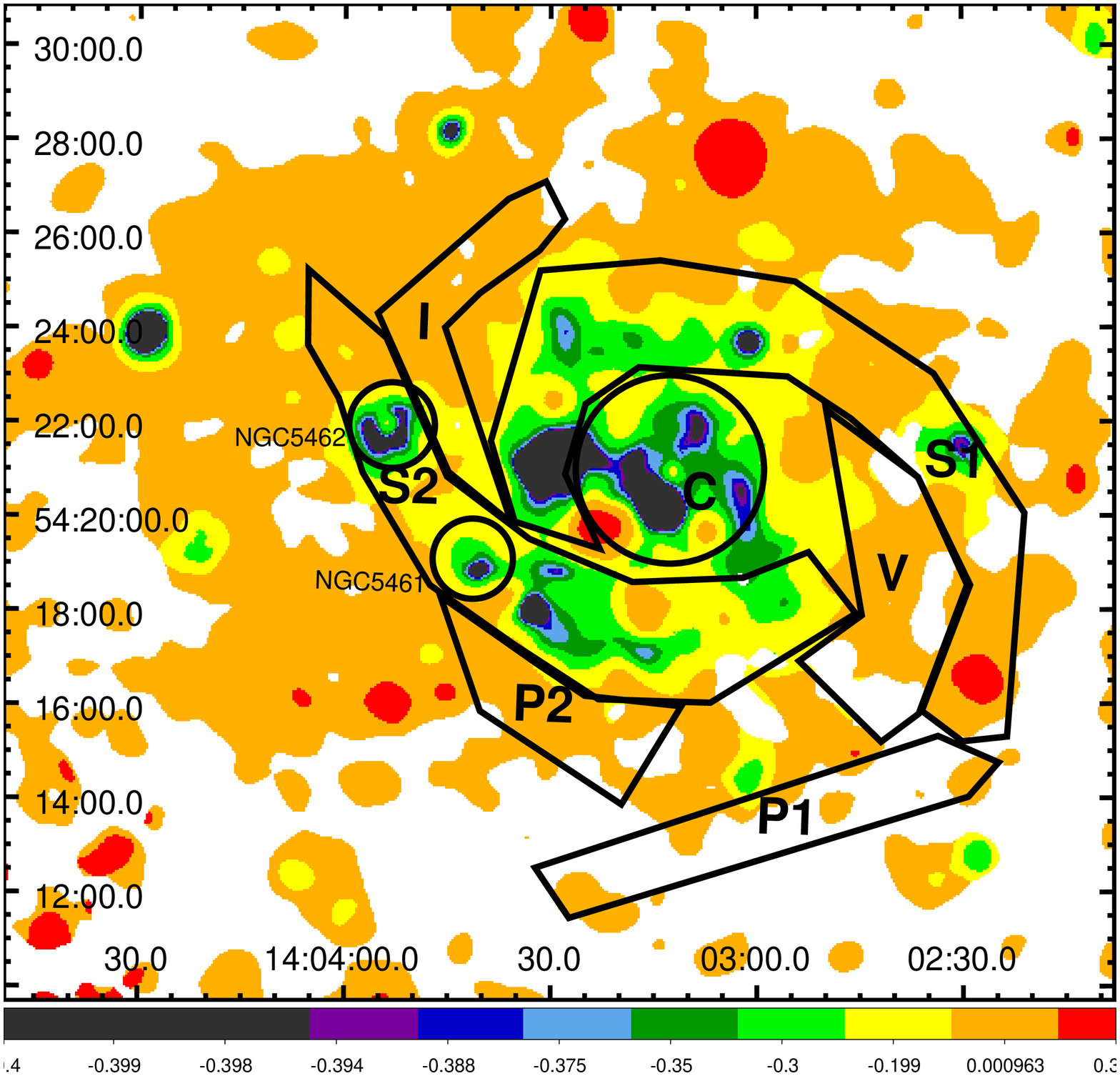}}
\caption{
        Regions used in the analysis of diffuse X-ray emission from M\,101 (see text for a detailed description)
        overlaid on the H$\alpha$ map (top left), the polarised radio intensity map from Fig.~\ref{6pi} (top right), 
        the soft X-ray emission map from Fig.~\ref{xmaps} (bottom left), 
        and the hardness ratio map from Fig.~\ref{m101hr} (bottom right).
        }
\label{m101xregs}
\end{figure*}

Contrary to the two galaxies mentioned above, M\,101 possesses only rudimentary `magnetic arms', that is, 
regions of enhanced polarised radio emission in areas
between the spiral stellar arms. Instead, we selected areas of significant polarised emission that are found just outside of the spiral arms. 
For a comparison, we also selected areas of the two most prominent spiral arms and the central 4$\arcmin$ of the galaxy that correspond to 
the inner part of the disk, having a diameter of about 8.6\,kpc at the assumed distance (see Table~\ref{astrdat}).
The most interesting region is region V, where only little \ion{H}{i} is visible and no H$\alpha$ or radio emission is present. 
However, significant X-ray emission in this region is visible. All regions are described in Table~\ref{names} and shown in Fig.~\ref{m101xregs}.

\begin{table}[ht]
\caption{\label{names}Regions of M\,101 used for the spectral analysis.}
\centering
\begin{tabular}{cl}
\hline\hline
Region name     & Region description                       \\
\hline
\vspace{5pt}
C               & central 4$\arcmin$ (8.6\,kpc) of M\,101 \\
\vspace{5pt}
P1               & southern polarised arm  \\
\vspace{5pt}
P2               & south-eastern polarised feature \\
\vspace{5pt}
I               & north-eastern polarised inter-arm region \\
\vspace{5pt}
S1               & western spiral arm \\
\vspace{5pt}
S2$_{\rm X}$     & eastern spiral arm \\
\vspace{5pt}
V                & `void' - area between the centre and S1 \\
\hline
\end{tabular}
\end{table}

We performed the spectral analysis for all of the regions. 
Because of the sensitivity of the data available for M\,101, for regions not associated with the spiral arms, it was not possible to fit a model 
with two thermal plasma components that would account for the emission from the disk and the halo. Instead, the fitted model to the spectra consisted 
of only one thermal plasma component to account for the entire diffuse emission from the hot gas. Therefore, we could obtain information about 
the combined emission from both the disk and the halo. The separation of both fractions, (i.e. the use of two thermal components), was possible 
for the models fitted to spectra from the spiral arms, as well as from the central region, which assured the sufficient number of net counts. 
All models have an additional power-law component that accounts for unresolved sources and the possible residual emission of sources excluded 
during the source detection procedure (see Sect.~\ref{background}). The thermal plasma was modelled with the use of the {\it mekal} model 
\citep{mewe85,kaastra92}. All models include the absorption component {\it wabs} to account for the influence of the neutral hydrogen in the halo 
of our Galaxy.

\begin{figure*}[ht]
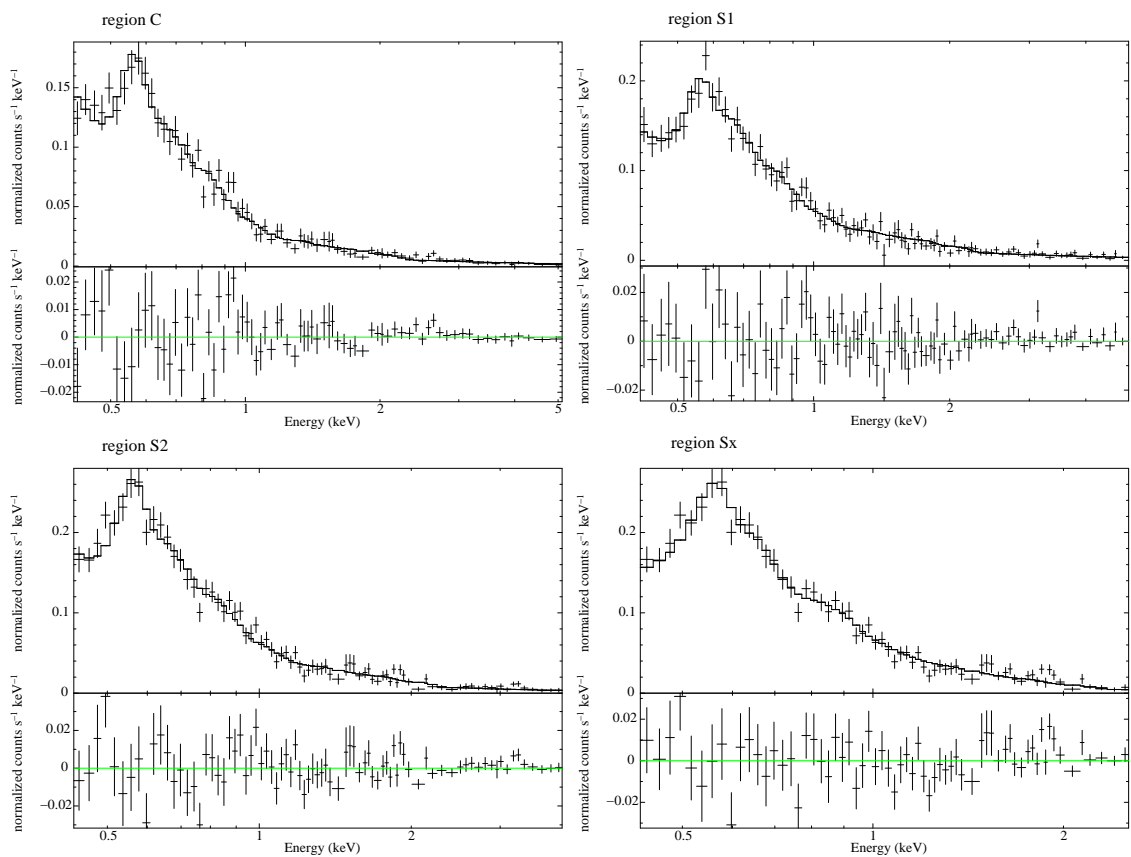

\begin{center}
\resizebox{0.4\hsize}{!}{\includegraphics[angle=-90]{m101_centre_normal.ps}}
\resizebox{0.4\hsize}{!}{\includegraphics[angle=-90]{m101_S1_normal.ps}}
\resizebox{0.4\hsize}{!}{\includegraphics[angle=-90]{m101_S2_normal.ps}}
\resizebox{0.4\hsize}{!}{\includegraphics[angle=-90]{m101_Sx_normal.ps}}
\end{center}
        \caption{Spectral model fits to the diffuse X-ray emission from star-forming regions in M\,101 (see Table~\ref{names} for description). 
Tables~\ref{m101xtabr} and \ref{m101xfr} present the parameters of the model fits and the derived fluxes, respectively.}
\label{m101models_sf}
\end{figure*}

\begin{figure*}[ht]
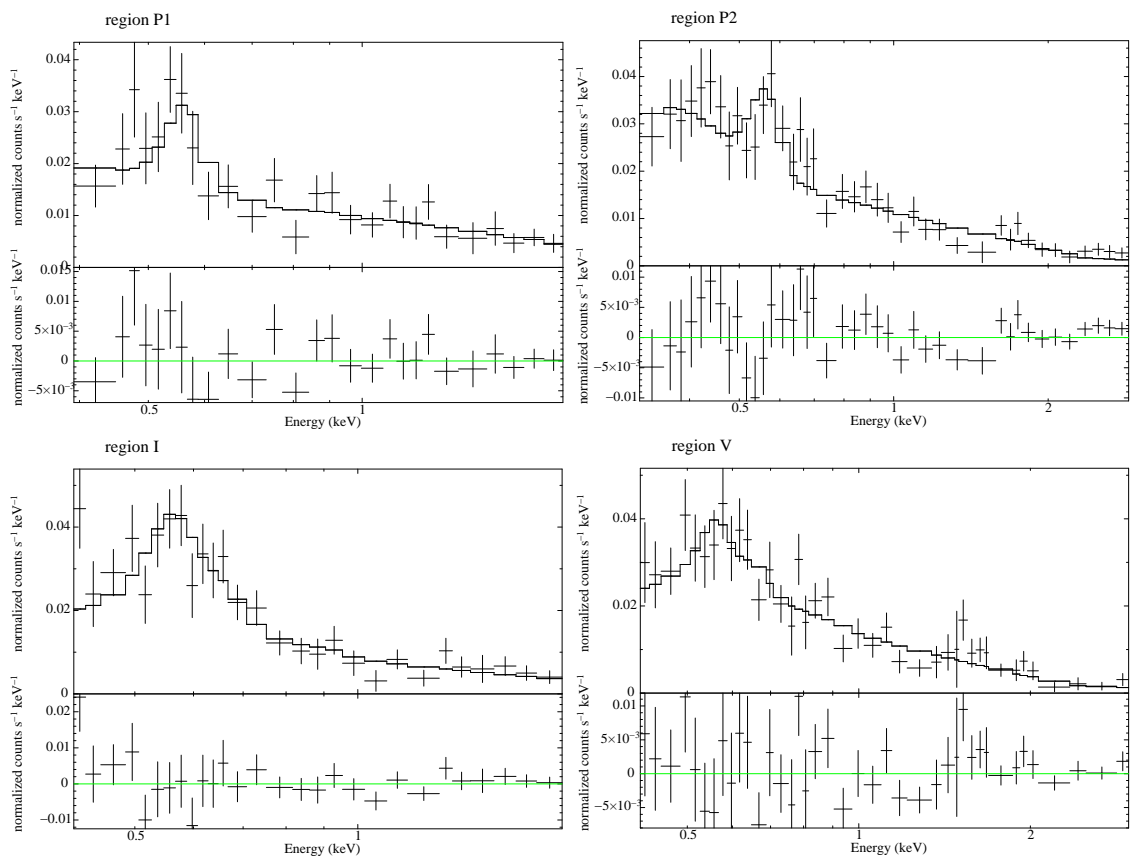

\begin{center}
\resizebox{0.4\hsize}{!}{\includegraphics[angle=-90]{m101_I1_normal.ps}}
\resizebox{0.4\hsize}{!}{\includegraphics[angle=-90]{m101_I2_normal.ps}}
\resizebox{0.4\hsize}{!}{\includegraphics[angle=-90]{m101_I3_normal.ps}}
\resizebox{0.4\hsize}{!}{\includegraphics[angle=-90]{m101_V1_normal.ps}}
\end{center}
        \caption{Spectral model fits to the diffuse X-ray emission from regions without star formation in M\,101 (see Table~\ref{names} for description).
Tables~\ref{m101xtabr} and \ref{m101xfr} present the parameters of the model fits and the derived fluxes, respectively.}
\label{m101models_null}
\end{figure*}

The models fitted to spectra from all of the regions are presented in Table~\ref{m101xtabr}, together with the derived parameters. The plots of all of the models are 
presented in Figs.~\ref{m101models_sf} and~\ref{m101models_null}, 
and the derived unabsorbed fluxes of all spectral components are shown in Table~\ref{m101xfr}. The spectrum 
from the spiral arm S2 could be fitted with two models, consisting of the same components but with different values of the parameters 
derived. In the first fit, temperatures similar to that found in the other two-temperature model fits were derived. The second fit (marked throughout 
this paper as S2$_{\rm X}$) resulted in higher temperatures (see Table~\ref{m101xtabr}) and a much steeper slope of the photon index of the power-law component. 
Nevertheless, the latter fits the data slightly better, as can be seen in the model curve and residuals for both models 
(bottom row of Fig.~\ref{m101models_sf}). 
We note here that such ambiguity was not found for the very similar spectrum of the S1 spiral arm region. It is possible that it was caused 
by the difference in sensitivities of both spectra (see Table~\ref{m101xtabr}), and the slight flattening around 0.9\,keV (likely attributed 
to the hotter thermal component) is not sufficiently pronounced in the spectrum of region S1 (Fig.~\ref{m101models_sf}).
On the other hand, the much higher photon index in the S2$_{\rm X}$ model fit might suggest that some of the (residual) emission 
from (unresolved) point sources in the spectrum from the S2 region, was accounted in the S2$_{\rm X}$ model fit for the hotter thermal component. 
We already discussed this phenomenon in \citet{wezgowiec16}. 
We argue that with the present data, it is difficult to find one of the models more plausible, especially when the uncertainties 
of the fitted model parameters are taken into account. 

Another important technical issue of our spectral fitting is related to reaching the lower limit for the {\sc mekal} temperature allowed by 
{\sc xspec12} when deriving uncertainties for the model fits to spectra from regions C, P2, S1, and S2 (low temperatures version). 
In the case of the spectrum from region C,  the fitted value of the temperature was also set to this lower limit. 
As a result, the lower uncertainties of the temperatures of the halo component for these regions are underestimated. We note here, however, 
that obtaining these lower limit values is unlikely to be related to the sensitivity or quality of the spectra, because the temperature values 
and their uncertainties are not correlated with the values of net counts or net counts per square arc-minute, as shown in Table~\ref{m101xtabr}. 
For example, the lower limit for the temperature was reached in the model fitted to the most sensitive spectrum in our analysis (region C). 
Because such low temperatures of the hot gas fit well the scenario of very low star-forming activity in the central disk of M\,101 (see Sect.~\ref{gasparams}), 
it is justified to claim that the derived low temperatures reflect real physical conditions of the hot gas, and 
especially given that the spectra of the CHANDRA data analysed by KS10 were also fitted with temperatures of $\sim$0.1 and $\sim$0.3\,keV.

The parameters of the fitted models suggest, however, an interesting possibility that the two-temperature models C, S1, and S2, refer only to 
the emission from the halo. In such a case, the faint disk emission would be detected indirectly in the low-sensitivity spectra as additional flux 
of one of the model components. The obtained temperatures of around 0.1\,keV and 0.3\,keV correspond well to typical temperatures derived 
from spectra of halos of edge-on galaxies \citep[e.g.][]{strickland04,tuellmann06}. 

As we mentioned before, for the most sensitive spectrum (spiral arm S2) it was possible to fit a different model (S2$_{\rm X}$), which yielded 
0.18$^{+0.01}_{-0.02}$ and 0.60$^{+0.19}_{-0.36}$\,keV for both thermal components, and a power-law component with a steeper photon index than 
in the low-temperature model.
In that model, the lower and the higher temperature referred to the halo and the disk, respectively. 
We note that such an interpretation does not contradict the findings of KS10. Although they obtained temperatures 
of $\sim$0.1 and $\sim$0.3\,keV for the global spectrum, the spectra of the emission from the two giant star-forming regions, NGC\,5461 and NGC\,5462, were best fitted 
with two thermal components with lower temperatures of $\sim$0.21 and $\sim$0.26\,keV, and higher temperatures of $\sim$0.60 and $\sim$0.72\,keV, respectively. 
Both these regions are found within our region S2 (and marked with labelled circles in Fig.~\ref{m101xregs}).

Therefore, we argue that the faint emission from the galactic disk of M\,101 might remain undetected in the analysed spectra, except for in the 
most sensitive one (spiral arm S2). For this one, it was also possible to fit a model that accounts for the hot disk component (owing to the bright emission 
from NGC\,5461 and NGC\,5462 star-forming regions), whose signs are visible as a `knee'-like feature around 0.9\,keV (see the bottom right panel 
in Fig.~\ref{m101models_sf}). Such a feature seems to be present also in the spectrum of the other spiral arm (S1, upper right panel 
in Fig.~\ref{m101models_sf}), as well as the central region (C, upper left panel in Fig.~\ref{m101models_sf}), but a fitting of the model with thermal components 
at temperatures around 0.2 and 0.6\,keV was not possible, likely due to lower sensitivity and the resulting larger spread of energy bins.

\begin{table*}[ht]
\caption{\label{m101xtabr}Model-fit parameters for the regions studied in M\,101.}
\centering
\begin{tabular}{clcccccc}
\hline\hline
Region& Model                      & kT$_1$                                  & kT$_2$                & Photon           &$\chi_{\rm red}^2$& Net   & Net cts \\
      & type                       & [keV]                                   & [keV]                 & Index                  &            & counts&/sqarcmin\\     \\
\hline
\vspace{5pt}
C     & wabs(mekal+mekal+power law)& 0.08\tablefootmark{a}$^{+0.01}_{-0.00}$ & 0.28$\pm$0.02              & 1.49$^{+0.21}_{-0.17}$ & 1.17       & 5152  & 447   \\
\vspace{5pt}
P1    & wabs(mekal+power law)      & 0.11$^{+0.04}_{-0.02}$                  & --                     & 1.59$^{+0.46}_{-0.47}$ & 0.87       & 1332  & 120   \\
\vspace{5pt}
P2    & wabs(mekal+power law)      & 0.09$^{+0.02}_{-0.01\tablefootmark{b}}$ & --                 & 1.95$^{+0.37}_{-0.35}$ & 1.14       & 1342  & 174   \\    
\vspace{5pt}
I     & wabs(mekal+power law)      & 0.17$^{+0.02}_{-0.03}$                  & --                     & 1.65$^{+1.04}_{-1.67}$ & 1.04       & 1322  & 176   \\
\vspace{5pt}
S1    & wabs(mekal+mekal+power law)& 0.09$^{+0.02}_{-0.01\tablefootmark{b}}$ & 0.26$\pm$0.03              & 1.44$^{+0.21}_{-0.20}$ & 0.99       & 6811  & 227   \\
\vspace{5pt}
S2    & wabs(mekal+mekal+power law)& 0.09$^{+0.03}_{-0.01\tablefootmark{b}}$ & 0.26$\pm$0.03              & 1.68$^{+0.28}_{-0.26}$ & 1.29       & 7545  & 303   \\
\vspace{5pt}
S2$_{\rm X}$& wabs(mekal+mekal+power law)& 0.18$^{+0.01}_{-0.02}$                    & 0.60$^{+0.19}_{-0.36}$& 2.42$^{+0.22}_{-0.27}$ & 1.22        & 7545  & 303   \\
\vspace{5pt}
V     & wabs(mekal+power law)      & 0.18$^{+0.02}_{-0.05}$                  & --                     & 1.98$^{+0.41}_{-0.48}$ & 1.08       & 1573  & 137   \\
\hline
\end{tabular}
\tablefoot{
        \tablefootmark{a}{ Value set at the lower limit of {\sc xspec12} (0.08\,keV).} 
        \tablefootmark{b}{ Lower uncertainty set at the lower limit of {\sc xspec12} (0.08\,keV)}
}
\end{table*}

\begin{table}[ht]
\caption{\label{m101xfr}Total (0.3 - 12 keV) unabsorbed fluxes in 10$^{-13}$erg\,cm$^{-2}$s$^{-1}$ for the modelled regions in M\,101.}
\centering
\begin{tabular}{cllll}
\hline\hline
Reg.            & mekal 1                & mekal 2                                      & power law                     & total   \\
\hline
\vspace{5pt}
C               & 3.4$^{+1.9}_{-1.5}$ (42) & 1.1$\pm$0.2 (14)                   & 3.6$^{+1.5}_{-1.1}$      (44) & 8.1$^{+3.5}_{-2.7}$     \\
\vspace{5pt}
P1              & 0.6$^{+2.7}_{-0.5}$ (19) & --                                 & 2.6$^{+2.9}_{-1.1}$      (81)   & 3.2$^{+5.5}_{-1.5}$   \\
\vspace{5pt}
P2              & 0.6$^{+1.2}_{-0.4}$ (33) & --                                 & 1.2$^{+0.6}_{-0.3}$      (67)   & 1.8$^{+1.8}_{-0.7}$   \\
\vspace{5pt}
I               & 0.3$^{+0.3}_{-0.2}$ (27) & --                                 & 0.8$^{+14.2}_{-0.5}$     (73) & 1.1$^{+14.4}_{-0.7}$    \\
\vspace{5pt}
S1              & 3.1$^{+5.7}_{-2.2}$ (25) & 1.8$\pm$0.5 (15)                   & 7.4$^{+3.6}_{-2.3}$      (60)   & 12.3$^{+9.8}_{-5.0}$  \\
\vspace{5pt}
S2              & 3.1$^{+9.0}_{-2.4}$ (34) & 1.5$\pm$0.4 (16)                   & 4.6$^{+2.6}_{-1.5}$      (50)   & 9.2$^{+12.1}_{-4.3}$  \\
\vspace{5pt}
        S2$_{\rm X}$            & 2.1$^{+0.4}_{-0.5}$ (34) & 0.3$^{+0.1}_{-0.2}$ (5)             & 3.8$^{+0.8}_{-0.5}$    (61)   & 6.2$^{+1.3}_{-1.2}$   \\
\vspace{5pt}
V               & 0.5$\pm$0.3         (19) & --                                 & 2.2$^{+1.8}_{-0.7}$      (81)   & 2.7$^{+2.1}_{-1.0}$   \\
\hline
\end{tabular}
\tablefoot{
Values in parentheses are fractions of the total flux (in percent) for a given component.
}
\end{table}

\section{Discussion}

\subsection{Parameters of the hot gas}
\label{gasparams}

To calculate  the physical parameters of the hot gas from the obtained model fits, assumptions about the emitting volumes need to be made.
M\,101 is known to present moderate star-forming activity \citep[2.33\,M$_{\sun}$/yr;][]{kennicutt11}, which is lower than that in M\,83 
\citep[3.20\,M$_{\sun}$/yr;][]{barnes14} and significantly lower than in NGC\,6946 \citep[7.12\,M$_{\sun}$/yr;][]{kennicutt11}. 
Because the extent of the X-ray halo of a galaxy depends on the star-forming activity in the underlying disk \citep[e.g.][]{strickland04b}, 
the moderate SFR of M\,101 suggests that an extent of the hot gaseous halo of 10\,kpc, as seen in many edge-on galaxies \citep[e.g.][]{tuellmann06}, 
is a justified assumption.

The volume assumptions and the model fit parameters allowed us to calculate number densities, gas masses, and thermal energies and their densities. 
This was done using the model of thermal cooling and ionisation equilibrium of \citet{nulsen84}: 
$$L_X=1.11\cdot \Lambda(T)\,n^2_e\,V\,\eta,$$ where $L_X$ is X-ray luminosity, $n_e$ is electron number density, $V$ is volume, $\eta$ is an unknown 
filling factor, and $\Lambda(T)$ is a cooling coefficient of the order of $10^{-22}\,{\rm erg}\,{\rm cm}^3\,{\rm s}^{-1}$ for temperatures 
of a few million K \citep{raymond76}. 

The obtained parameters of the hot gas are presented for both of the thermal components separately (Tables~\ref{halo} and~\ref{disk}). 
In Sect.~\ref{spectra}, we argue that the two thermal component models describe the emission mainly from the hot gas in the halo. 
It was possible to tract the contribution 
from the disk gas only in the case of the most sensitive spectrum from the spiral arm S2. In the alternative model fitted 
to this spectrum (S2$_{\rm X}$), the flux calculated for this disk component amounts to just 5\% of the total model flux. This of course does not mean 
that this weak emission from the hot gas in the disk is not observed in other spectra related to the star forming arms, 
but rather that it cannot be separated from the significantly more pronounced 
halo emission. For the single thermal component model fits, the straightforward assumption is that the entire emission is described with only one temperature, 
which consists of contributions from both the disk and halo emission at an unknown ratio. Consequently, to be able to compare the properties of the hot gas
in all of the regions, we need to calculate the global values from the two thermal components model fits (i.e. for regions C, S1, S2, and S2$_{\rm X}$). 
The temperatures and number densities of both of the thermal components were averaged and the gas masses and thermal energies 
were simply added. As a result we obtained the new set of global values, which are presented in Table~\ref{totalhot}, together with the values 
calculated for regions for which single-thermal component models were fitted.

\begin{table*}[ht]
\caption{\label{halo}Derived parameters of the hot gas in the first thermal component of M\,101.}
\centering
\begin{tabular}{ccccc}
\hline\hline
Region& n$^{1st}\eta^{-0.5}$   & M$^{1st}_{gas}\eta^{0.5}$ & E$^{1st}_{th}\eta^{0.5}$ & $\epsilon^{1st}_{th}\eta^{-0.5}$    \\
\vspace{5pt}&[10$^{-3}$cm$^{-3}$]& [10$^6$M$_\odot$]       & [10$^{54}$\,erg]         & [10$^{-12}$\,erg\,cm$^{-3}$]        \\
\hline
\vspace{5pt}
C     & 6.9$^{+0.4}_{-1.7}$    & 92$^{+5}_{-23}$           & 21$^{+4}_{-5}$           & 1.33$^{+0.25}_{-0.33}$               \\
\vspace{5pt}
P1    & 1.7$^{+1.5}_{-0.7}$    & 25$^{+21}_{-10}$          & 8$^{+12}_{-4}$           & 0.46$^{+0.70}_{-0.23}$               \\
\vspace{5pt}
P2    & 2.9$^{+1.0}_{-1.1}$    &  28$\pm$10                & 7$^{+5}_{-3}$            & 0.62$^{+0.40}_{-0.27}$               \\
\vspace{5pt}
I     & 1.1$^{+0.3}_{-0.2}$    & 10$^{+3}_{-2}$            & 5$\pm$2                  & 0.45$^{+0.19}_{-0.15}$               \\
\vspace{5pt}
S1    & 3.7$^{+0.8}_{-1.5}$    & 129$^{+26}_{-53}$         & 33$\pm$16                & 0.81$\pm$0.38                        \\
\vspace{5pt}
S2    & 3.8$^{+1.4}_{-1.7}$    & 108$^{+39}_{-48}$         & 28$^{+23}_{-14}$         & 0.82$^{+0.66}_{-0.41}$               \\
\vspace{5pt}
S2$_{\rm X}$& 1.6$\pm$0.1            & 45$\pm$3                  & 23$^{+3}_{-4}$           & 0.67$^{+0.09}_{-0.12}$         \\
\vspace{5pt}
V     & 1.1$\pm$0.1            & 15$\pm$3                  &  8$^{+2}_{-3}$           & 0.45$^{+0.14}_{-0.20}$               \\
\hline
\end{tabular}
\tablefoot{
Columns are the region name, electron number density, total gas mass, total thermal energy, and thermal energy density.
$\eta$ is the unknown volume filling factor.
}
\end{table*}

\begin{table*}[ht]
\caption{\label{disk}Derived parameters of the hot gas in the second thermal component of M\,101.}
\centering
\begin{tabular}{ccccc}
\hline\hline
Region& n$^{2nd}\eta^{-0.5}$   & M$^{2nd}_{gas}\eta^{0.5}$ & E$^{2nd}_{th}\eta^{0.5}$ & $\epsilon^{2nd}_{th}\eta^{-0.5}$ \\
\vspace{5pt}
      &[10$^{-3}$cm$^{-3}$]    & [10$^6$M$_\odot$]         & [10$^{54}$\,erg]         & [10$^{-12}$\,erg\,cm$^{-3}$]     \\
\hline
\vspace{5pt}
C     & 4.5$^{+0.4}_{-0.3}$    & 6.0$^{+0.5}_{-0.4}$       & 4.8$^{+0.8}_{-0.7}$      & 3.0$^{+0.5}_{-0.4}$               \\
\vspace{5pt}
S1    & 3.7$^{+0.3}_{-0.5}$    & 12.8$^{+1.3}_{-1.6}$      & 9.5$^{+2.2}_{-2.1}$      & 2.3$\pm$0.5                       \\
\vspace{5pt}
S2    & 3.7$^{+0.4}_{-0.5}$    & 10.6$^{+1.2}_{-1.4}$      & 7.9$^{+1.9}_{-1.8}$      & 2.3$^{+0.6}_{-0.5}$              \\
\vspace{5pt}
S2$_{\rm X}$& 1.4$^{+0.4}_{-0.5}$    & 4.0$^{+1.1}_{-1.5}$       & 6.9$^{+4.6}_{-5.2}$      & 2.0$^{+1.4}_{-1.5}$        \\
\hline
\end{tabular}
\tablefoot{
Columns are the region name, electron number density, total gas mass, total thermal energy, and thermal energy density. 
$\eta$ is the unknown volume filling factor.
}
\end{table*}

\begin{table*}[ht]
\caption{\label{totalhot} Global parameters of the hot gas in the spectral regions of M\,101.}
\centering
\begin{tabular}{cccccc}
\hline\hline
Reg.& kT                      & n$^{avg}\eta^{-0.5}$    & M$^{tot}_{gas}\eta^{0.5}$     & E$^{tot}_{th}\eta^{0.5}$ & $\epsilon^{tot}_{th}\eta^{-0.5}$\\
\vspace{5pt}
    & [keV]                   & [10$^{-3}$cm$^{-3}$]    & [10$^6$M$_\odot$]             & [10$^{54}$\,erg]         & [10$^{-12}$                     \\
\vspace{5pt}
    &                         &                         &                               &                          & erg\,cm$^{-3}$]                 \\
\hline
\vspace{5pt}
C     & 0.09$^{+0.01}_{-0.00}$& 6.70$^{+0.38}_{-1.61}$  & 98$^{+6}_{-24}$               & 26$^{+5}_{-6}$           & 1.48$^{+0.27}_{-0.34}$          \\
\vspace{5pt}
P1    & 0.11$^{+0.04}_{-0.02}$& 1.73$^{+1.46}_{-0.68}$  & 25$^{+21}_{-10}$              & 8$^{+12}_{-4}$           & 0.46$^{+0.70}_{-0.23}$          \\
\vspace{5pt}
P2    & 0.09$^{+0.02}_{-0.01}$& 2.86$^{+1.00}_{-1.06}$  & 28$\pm$10                     & 7$^{+5}_{-3}$            & 0.62$^{+0.40}_{-0.27}$          \\
\vspace{5pt}
I     & 0.17$^{+0.02}_{-0.03}$& 1.10$^{+0.30}_{-0.20}$  & 10$^{+3}_{-2}$                & 5$\pm$2                  & 0.45$^{+0.19}_{-0.15}$          \\
\vspace{5pt}
S1    & 0.11$^{+0.02}_{-0.00}$& 3.72$^{+0.73}_{-1.42}$  & 141$^{+28}_{-54}$             & 43$\pm$18                & 0.94$\pm$0.40                   \\
\vspace{5pt}
S2    & 0.10$^{+0.03}_{-0.00}$& 3.76$^{+1.27}_{-1.56}$  & 119$^{+40}_{-49}$             & 36$^{+25}_{-16}$         & 0.95$^{+0.65}_{-0.42}$          \\
\vspace{5pt}
S2$_{\rm X}$& 0.21$^{+0.04}_{-0.05}$& 1.54$^{+0.13}_{-0.15}$    & 49$^{+4}_{-5}$                & 30$^{+8}_{-9}$           & 0.79$^{+0.20}_{-0.24}$          \\
\vspace{5pt}
V     & 0.18$^{+0.02}_{-0.05}$& 1.05$^{+0.18}_{-0.24}$  & 15$\pm$3                      &  8$^{+2}_{-3}$           & 0.45$^{+0.14}_{-0.20}$          \\
\hline
\end{tabular}
\tablefoot{
Columns are the region name, normalisation-weighted average temperature, electron number density, total gas mass, total thermal energy, 
        and thermal energy density. $\eta$ is the volume filling factor.
}
\end{table*}

It is not surprising that the highest average temperature of the hot gas is derived from the S2$_{\rm X}$ model fit, which accounts for the emission 
from the spiral arm S2 (especially the two giant star-forming regions) and the halo above it. 
However, this value agrees (within errors) with temperatures derived for the inter-arm polarised region I, 
where no significant star formation is visible, as well as temperatures derived for the `void' region V, 
in which no emission other than that visible in the X-ray map is present. 
Because, as we argued before, most of the observed X-ray emission comes from the halo of M\,101, we expect that in these two regions the temperature 
of the gas halo is comparable to that derived from the S2$_{\rm X}$ model. While the latter can easily be due to the underlying massive 
star-forming arm, in the case of regions I and V, a different explanation is needed. 

Of course, there is a possibility that the hot gas found in these areas originated in the nearby regions of star formation, for example at the end 
of the spiral arm S1 or in NGC\,5462 (Fig.~\ref{m101xregs}). Another possibility would be to associate these higher temperatures with lower gas densities 
(Table~\ref{totalhot}), but the large uncertainties make it difficult, especially given that the gas density derived 
from the S2$_{\rm X}$ model is comparable to that in the P1 region, another area without significant star formation, and hosting 
the hot gas at a temperature as low as 0.11$^{+0.04}_{-0.02}$\,keV. 

Interestingly, the analysis of the FUV data performed by KS10 suggests that the flow of the hot gas due to rotation of the galaxy 
should be mainly along the spiral arms, and therefore a detection of this gas is unlikely at larger distances from the arms.
The difference between the distribution of the hot gas in the maps derived from CHANDRA and XMM-Newton data can be a result of a much higher sensitivity 
to diffuse emission, but also lower resolution of the latter. We note here that because of the angular size of M\,101, the resolution 
difference can be neglected, contrary to the sensitivity to faint large-scale structures. Therefore, the detection of the diffuse X-ray emission 
from the inter-arm regions in our maps could be further evidence that the hot gas was not transported from the spiral arms, but rather heated or 
reheated on site.

Nevertheless, the comparison of gas densities and temperatures does not take into account the unknown filling factor 
($\eta$), which can and likely does vary across the galaxy. To remove its influence on our results and following our earlier studies 
\citep{wezgowiec16,wezgowiec20}, for each region we calculated the energy per particle $E_p$, which is the ratio of the thermal energy density and the electron 
density of the hot gas. Because both parameters show the same dependence on the filling factor, the calculated energies per particle can be more easily 
related to gas temperatures, and also compared with the energy densities of the magnetic fields. This is discussed in Sect.~\ref{hotfields}.

Last, but not least, we note that the derived densities of the hot gas in M\,101 (Table~\ref{totalhot}) somewhat exclude the possibility 
that cosmic-ray streaming, mentioned in Sect.~\ref{intro}, could be the main heating mechanism. Assuming a filling factor of one, 
gas at densities of a few 10$^{-3}$cm$^{-3}$ would be heated to around 10$^6$\,K, 
which is comparable to the lowest temperature that we derived from the fit (0.08\,keV). Lower values of the filling factor, down to 0.23 for the gas 
in the disk \citep{deavillez00}, will result in higher gas densities and consequently even lower temperatures. 

\subsection{Magnetic fields in M\,101}
\label{bfields}

For all of the regions for which the spectral analysis of the X-ray emission was performed, 
we measured fluxes in the total and polarised radio intensity maps 
(Figs.~\ref{6tp} and~\ref{6pi}). Then, using the energy equipartition revised formula developed by \citet{beck05}, we calculated strengths 
of the total magnetic field, as well as its ordered component and its total energy density. Because the information about the magnetic fields 
can be obtained only from the non-thermal component of the radio emission, for our calculations we needed to estimate the contribution of this component 
in the observed total radio emission. To do that, we used the findings of \citet{berkhuijsen16}, who estimated an average thermal fraction 
to be $f_{th}$ = 0.45. Because this was based on low-resolution observations and the nature of our analysis is to compare regions of star formation 
with those residing between the star-forming spiral arms, we decided to use a thermal fraction of $f_{th}$ = 0.5 for regions with significant 
star formation (C, S1, and S2) and considered the radio emission from ``inter-arm regions'' P1, P2, I, and V to be purely non-thermal. 
Such assumptions are likely conservative and the derived non-thermal intensities could be considered as upper limits. Therefore, to minimise this effect, 
in our error calculations we used the derived non-thermal intensities decreased by 25\%. We also used the non-thermal spectral index of the radio emission
from M\,101 of $\alpha_{nth}$ = 0.92, as derived by \citet{grave90} and used by \citet{berkhuijsen16}, as well as assumed a proton-to-electron ratio of 100. 
For the path-length we used a disk thickness of 1\,kpc. We note here that the uncertainties in the proton-to-electron ratio and the path-length 
result in systematic errors, affecting measurements for all of the regions in the same manner. 
Therefore, in our error calculations we included only the lower non-thermal 
intensities mentioned before. The calculated properties of the magnetic fields in the studied regions of M\,101 are presented in Table~\ref{magparams}.

\begin{table*}[ht]
\caption{\label{magparams} Properties of the magnetic fields in M\,101.}
\centering
\begin{tabular}{cccccc}
\hline\hline
Region          &S$_{synch}$    &p$_{synch}$    &B$_{tot}$              &$\epsilon_B$                         &B$_{ord}$    \\
\vspace{5pt}    &[mJy/beam]     &[\%]           &[$\mu$G]               &[10$^{-12}$\,erg\,cm$^{-3}$]         &[$\mu$G]     \\
\hline
\vspace{5pt}
C               & 0.45          & 14.9          & 10.7$\pm$0.8          & 4.6$\pm$0.7                     & 4.1$\pm$0.4 \\
\vspace{5pt}
P1              & 0.13          & 37.5          &  7.5$\pm$0.7          & 2.2$\pm$0.3                     & 4.9$\pm$0.4 \\
\vspace{5pt}
P2              & 0.26          & 20.8          &  9.2$\pm$0.7          & 3.4$\pm$0.5                     & 4.3$\pm$0.4 \\
\vspace{5pt}
I               & 0.18          & 32.3          &  8.2$\pm$0.8          & 2.7$\pm$0.5                     & 4.9$\pm$0.3 \\
\vspace{5pt}
S1              & 0.28          & 17.0          &  9.4$\pm$0.7          & 3.5$\pm$0.5             & 3.9$\pm$0.4 \\
\vspace{5pt}
S2              & 0.41          & 17.9          & 10.4$\pm$0.9          & 4.3$\pm$0.7             & 4.4$\pm$0.4 \\
\vspace{5pt}
V               & 0.19          &  5.5          &  8.8$\pm$0.7          & 3.1$\pm$0.5             & 2.0$\pm$0.1 \\
\hline
\end{tabular}
\tablefoot{
Columns are the region name, non-thermal radio flux, degree of polarisation, total magnetic field strength, magnetic field energy, 
        and ordered magnetic field strength.
}
\end{table*}

The derived strengths of the magnetic field are between 9 and 11\,$\mu$G for most of the regions, 
which corresponds well with the findings by \citet{berkhuijsen16}. 
Lower values were obtained for three regions: P1, I, and V, which do not correspond to regions of star formation. 
Consequently, these three areas show the lowest energy densities of the magnetic field. Two of them, regions P1 and I, 
show at the same time the highest strengths of the ordered component of the magnetic field, reaching 
as high as 4.9\,$\mu$G. This suggests a significant ordering of the magnetic field in M\,101. 
The third one of these regions (V) shows a comparable strength of the total magnetic field, but the strength of the ordered 
component is significantly lower than for any other regions within the galactic disk. A straightforward explanation 
is the fact that region V looks like a large hole in the map of the polarised radio emission from M\,101 (Fig.~\ref{6pi}).
This hole is accompanied by a lack of any significant star formation, with a prominent depression visible in the \ion{H}{i} map.
If we assume that the radio emission has been depolarised by Faraday rotation (e.g. by vertical magnetic fields) in this area, 
this would need to  
be a strong effect, as the degree of polarisation is lower by a factor of three or more than in other regions. Such significant
depolarisation would require rotation measures of |RM| > 300\,rad/${\rm m}^2$, while the observed ones are much smaller, 
as shown by \citet{berkhuijsen16}. Therefore, the lack of any substantial polarised emission from the ordered magnetic field seems to be real.

The two spiral arms and the central region of the galaxy show the highest strengths, and 
thus the highest energy densities of the magnetic field. 
Although this seems obvious, given the star-forming activity in these regions, 
significant emission can be seen in the polarisation map. This is reflected in the strengths of the 
ordered component of the magnetic field, which are comparable to those of the remaining regions. 
The eastern spiral arm S2 shows a slightly stronger magnetic field and with a higher energy density than 
the western spiral arm S1. Because this arm looks distorted, likely due to intra-group interactions, 
this increase could be explained by the gravitation-induced enhancement 
of the star formation and consequently the amplification of the magnetic field. The tidal interactions could also simply 
stretch the magnetic fields, leading to their amplification and the enhancement of the ordered component, which is also observed in this arm. 

\subsection{Hot gas and magnetic fields in M\,101}
\label{hotfields}

As mentioned in Sect.~\ref{gasparams}, the X-ray observations of M\,101 provide information
about the physical properties of the hot gas both in the galactic disk and the halo.
Although the observed non-thermal radio emission also comes from the entire galaxy (being integrated along the line of sight),
we do not have any direct information about the halo magnetic fields, (i.e. their structure and strengths),
because in this paper the radio data at only one frequency are presented. Therefore, being unable to disentangle
the radio emission from the disk from that from the halo, we needed to calculate the global properties of the hot gas,
as presented in Sect.~\ref{gasparams}, including the energies per particle, which provide direct information about
the energy stored in the hot gas. Consequently, we could compare these energies per particle with the energy densities
of the magnetic fields for all of the regions of M\,101 that are analysed in this paper.

\begin{figure}[ht]
\resizebox{\hsize}{!}{\includegraphics[clip,angle=-90]{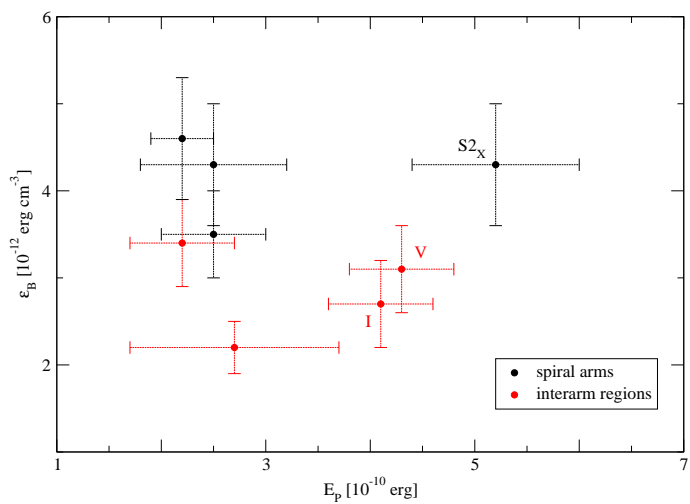}}
        \caption{Relation between energy per particle (${\rm E}_{\rm P}$) and energy density of the magnetic field (${\rm\epsilon}_{\rm B}$
        in the regions of M\,101.
        }
\label{epmag}
\end{figure}

\begin{figure}[ht]
\resizebox{\hsize}{!}{\includegraphics[clip,angle=-90]{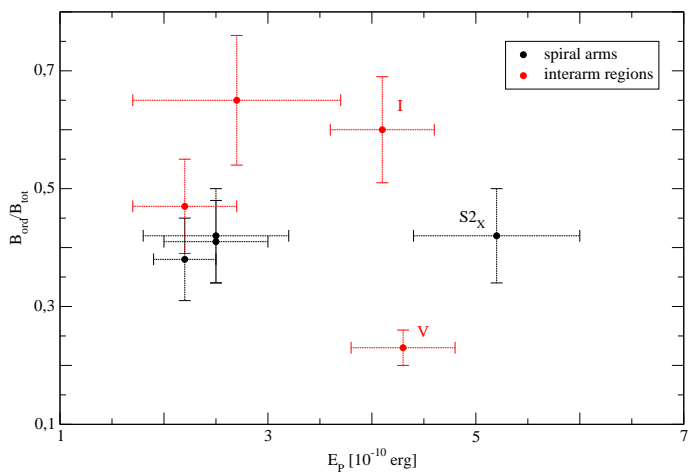}}
        \caption{Relation between energy per particle (${\rm E}_{\rm P}$) and degree of order (${\rm B}_{\rm ord}/{\rm B}_{\rm tot}$)
        of the magnetic field in the regions of M\,101.
        }
\label{epbord}
\end{figure}

In Table~\ref{particles}, the energies per particle together with energy densities of the magnetic field are presented
and their relation is shown in Fig.~\ref{epmag}.
The highest energies per particle are calculated for the polarised inter-arm region I (magenta), S2$_{\rm X}$
(the higher temperature model for spectrum of the spiral arm S2), and the inter-arm region V (blue). The corresponding energy densities
of the magnetic field are lower for I and V, when compared to other regions, but among the highest in the case of S2$_{\rm X}$.
Fig.~\ref{epbord} presents the relation of the energies per particle and the degree of order of the magnetic field, defined here
as the ratio of the strengths of the ordered and the total magnetic fields.
The values that are clearly above the average are for one of the two polarised regions (P1, red dot) and for I.
The lowest degree of order of the magnetic field is found, as discussed before, in region V.

As we can see, all three conditions required to suggest the magnetic reconnection is a mechanism of the ISM heating (Sect.~\ref{intro}) are
met only in region I, which is in fact the only magnetic arm in M\,101. While the strong and moderately ordered 
magnetic fields in the spiral arm S2, together with the highest temperature of the hot gas (derived from S2$_{\rm X}$ model), 
are likely due to the tidal interactions discussed in Sect.~\ref{bfields}, the properties of the hot gas 
and magnetic field in region V need a quite different explanation. Interestingly, the properties of the hot gas, that is, 
temperatures, number densities, and thermal energy densities are in fact identical in regions I and V.
Furthermore, the total strengths of the magnetic fields, as well as their energy densities are also very similar, with the values for region V
being slightly higher, though still within the derived uncertainties. This makes the degree of order of the magnetic field the only property that
is quite different between both regions. Because of the significant extent of region V (roughly 5\,$\times$10\,kpc) and the absence of star formation,
only little turbulence is expected in this area, which in turn would prevent the formation of ordered magnetic fields.
This, however, would not explain the higher temperature of the hot gas in this area, when compared to the other 
regions,  especially the neighbouring regions,  of M\,101. As the hot gas in region V is expected to originate in the nearby 
star-forming regions, such as the one visible at the end of the S1 spiral arm (Fig.~\ref{m101xregs}), its temperature should be comparable to that 
of the gas in regions P1 or P2. In fact, we note here that the sensitivity of the spectrum from region V is low and it is 
likely that the real temperature of the hot gas in this region is indeed comparable to the temperatures in P1 and P2, as the uncertainties suggest. 
Therefore, more sensitive X-ray data from region V are needed to verify the temperature of the hot gas. 

If the present temperature was confirmed, the existence of the hotter gas in region V could be, to some extent, also due to the 
effects of magnetic reconnection. These would operate on the vertical component of the magnetic field in this area. Although 
the Faraday rotation measures obtained by \citet{berkhuijsen16} are not sufficient to support the depolarisation scenario, they are higher 
in this region than in the other parts of the galaxy, which would suggest the existence of (weaker) vertical magnetic fields. In such a case, 
the increase in the degree of order of this vertical component would not be visible in our map of the polarised emission 
because this emission is sensitive only to the field components in the sky plane.

\begin{table}[ht]
\caption{\label{particles}Thermal energy per particle and energy densities of the magnetic field in the spectral regions of M\,101.}
\centering
\begin{tabular}{ccc}
\hline\hline
Region      & $E_p$               & $\epsilon_{B}$\\
\vspace{5pt}
            & [10$^{-10}$\,erg]   & [10$^{-12}$\,erg\,cm$^{-3}$]\\
\hline
\vspace{5pt}
C           & 2.2$\pm$0.3         & 4.6$\pm$0.7 \\
\vspace{5pt}
P1          & 2.7$\pm$1           & 2.2$\pm$0.3 \\
\vspace{5pt}
P2          & 2.2$\pm$0.5         & 3.4$\pm$0.5 \\
\vspace{5pt}
I           & 4.1$\pm$0.5         & 2.7$\pm$0.5 \\
\vspace{5pt}
S1          & 2.5$\pm$0.5         & 3.5$\pm$0.5 \\
\vspace{5pt}
S2          & 2.5$\pm$0.7         & 4.3$\pm$0.7 \\
\vspace{5pt}
S2$_{\rm X}$& 5.2$\pm$0.8         & 4.3$\pm$0.7  \\
\vspace{5pt}
V           & 4.3$\pm$0.5         & 3.1$\pm$0.5 \\
\hline
\end{tabular}
\end{table}

\subsection{Reconnection heating}
\label{reconheat}

As already suggested in Sect.~\ref{intro}, in order to propose magnetic reconnection as the source of gas heating, three 
conditions need to be fulfilled simultaneously. Apart from the increase in the temperature of the hot gas, a lower energy density 
of the magnetic field should be observed, as some of the kinetic and magnetic energy of turbulence is being dissipated and converted 
to thermal energy in the surrounding medium. Additionally, a decrease in the turbulent energy of the magnetic field would result 
in a lower ratio of total and ordered magnetic fields, and thus their higher level of magnetic field ordering. All these properties of the hot gas and 
the magnetic fields seem to be present in region I (the only magnetic arm), as already mentioned in the previous section. 
Still, it is important to ask if the increase in temperature of the hot gas due to reconnection heating can be detected,
and if the required energy of the magnetic field is consistent with the values derived from our observations. 
We discuss these problems and present our estimates in the following sections.

\subsubsection{Gas heating}
\label{heating}

The chances of detecting reconnection heating in galaxies
depend on where and when the energy of dissipated magnetic fields is released. The essential difficulty stems from the fact
that most of the supernova explosion energy
(canonical value of 10$^{51}$\,erg) is converted to thermal energy during the shock passing through
the ISM. Shock waves in the ISM heat up the post-shock gas to 10$^6$ - 10$^7$\,K. Expansion of the over-pressured bubbles, 
especially those related to multiple supernova explosions in star-forming regions, drives turbulence. According to \citet{maclow04}, 
about $\sim$10\% of supernova energy is converted to turbulent
energy, which drives the turbulence in the ISM. Another 10-50\% of supernova energy is spent on acceleration of cosmic rays \citep[e.g.][]{grenier15}.
The remaining part (40-80\%) of the explosion energy is deposited in the hot-ISM phase. Hot gas together with cosmic rays drive galactic outflows.
The hot bubbles expand and cool down adiabatically, while the hot gas cools down via radiative processes.
Turbulent cascade transports the kinetic and magnetic energies of chaotic flows to smaller and smaller scales. Assuming that magnetic energy
constitutes about half of the turbulent energy density, 
we note that $\sim$5\% of the supernova explosion energy is absorbed in the amplification of random magnetic fields.
In a turbulence decay timescale, which is of the order of eddy turnover time, the chaotic fluctuations reach dissipation scales, 
where turbulent reconnection \citep{lazarian20} operates and dissipates magnetic fields, converting their energy into thermal energy. 

The above estimations lead to the conclusion  
that reconnection heating can contribute $\sim$10\% of energy of the hot-ISM phase. As a result, it seems that reconnection heating
is sub-dominant and difficult to detect if both heating processes act simultaneously.
However, it is possible that reconnection heating is delayed and/or displaced from star-forming regions, where the hot gas phase 
is most efficiently produced.

To identify the plausible time lags and space shifts of magnetic field dissipation in turbulence, we note that supernovae produce hot gas during
the Sedov phase, when a part of the supernova energy is converted to kinetic energy of the expansion. This leads to motions of gas on scales of the order
of supernova or super-bubble (SB) scales, which range from $\sim$\,50\,pc to 300\,pc. Parker instability, due to buoyancy of magnetic fields and cosmic rays,
is considered as another driver of interstellar turbulence on even larger scales of 0.5 - 1\,kpc.
These scales can be identified as energy injection scales (outer scale) for ISM turbulence, corresponding to velocities at the outer scale to 10\,km\,s$^{-1}$
and 50\,km\,s$^{-1}$, respectively \citep[see][]{shukurov19}. Knowing the scales ($L$) and velocities ($v$), we can estimate the time 
lag between the energy injection on the outer scale
of turbulence and the dissipation of kinetic and magnetic energies at the dissipation scales. 
With the above values, the full cascade and the associated reconnection takes place
in about 5 to 10\,Myr in both the disk and the halo. In that time, the disk interstellar gas, 
orbiting with a velocity of 200\,km\,s$^{-1}$ ($\simeq$200\,pc\,Myr$^{-1}$),
covers the azimuthal distance of $\Delta L_\varphi \sim$1-2\,kpc.
Inside the co-rotation radius, the difference between the ISM gas speed and the pattern speed can lead to a significant
displacement of the turbulent ISM with respect to the spiral arms. The plausible scenario is that the turbulence driven in the spiral arms 
gradually decays with the distance from the arms. The effects of magnetic reconnection result in the ordering of the magnetic field 
structure, and these should both be noticeable in the wakes of the past star forming regions, or in general at distances 
of the order of $\sim$1--2\,kpc. If this scenario is valid, the azimuthal distance $\Delta L_\varphi$ denotes the displacement 
of the magnetic arms with respect to the spiral arms.

On the other hand, supernova shock-heated gas expands rapidly and drives both turbulence and mostly vertical outflows.
The rapid expansion of the over-pressured gas leads to adiabatic cooling, which acts together with the radiative cooling processes.
Between the initial temperature of shock-heated gas of $\sim$10$^7$\,K and temperatures around 10$^6$\,K, the adiabatic cooling dominates
over radiative processes. Due to the bump in the cooling curve \citep{dalgarno72},
where the cooling rates increase from $\sim$2$\times$10$^{-23}$erg\,cm$^{3}$\,s$^{-1}$ at 10$^6$\,K to $\sim$10$^{-21}$\,erg\,cm$^{3}$\,s$^{-1}$ at around
5$\times$10$^5$\,K, the radiative cooling processes become significantly more efficient below 10$^6$\,K. Then, the cooling times
may fall below 10\,Myr for the gas at number densities of a few 10$^{-3}$\,cm$^{-3}$ \citep[see eq. 38 and Fig. 4. in][]{hanasz98}.
This means that supernova-driven outflows have a good chance to cool down in a period shorter 
than or comparable to the period of turbulence dissipation and magnetic reconnection. Therefore, when conditions are favourable, 
the shock-heated gas may already be significantly cooled down, while the reconnection heating still operates.
The outflow velocities of the post-shock gas are of the order of the escape velocity from the disk of $\sim$300\,km/s or more, which
follows the thermal expansion acting together with the outflow acceleration by cosmic ray pressure gradient.
While thermal expansion might fail due to the rapid cooling of the hot gas fraction, cosmic rays lose their energy much less efficiently
and can therefore serve as a wind driver even without the active support from the hot gas fraction \citep{hanasz13}.

We estimate that in 5-10\,Myr, the outflow of the hot gas propagating at velocities of 300\,km/s covers a vertical distance 
of $\Delta L_z\sim$1.5-3\,kpc and at the same time an azimuthal
distance of $\Delta L_\varphi$$\sim$1-2\,kpc, because the out-flowing gas continues its orbital movement. 
This implies that the effects of magnetic reconnection in the hot gas
can potentially be detectable at vertical distances $\pm\Delta L_z$ above and below the disk, with the azimuthal displacement of $\Delta L_\varphi$.
This means that radiative losses of the reconnection-heated gas might be located above and below the inter-arm regions.

\subsubsection{Dissipation of the magnetic field}
\label{turbdis}

The total energy densities of the magnetic fields in the studied regions of M\,101 are roughly 2-5$\cdot$10$^{-12}$\,erg\,cm$^{-3}$ (Table~\ref{magparams}).
Because the strengths of the ordered magnetic fields are two to three times lower than those of the total magnetic fields, we can assume 
that about 80\% of the total energy is stored in the turbulent component, which gives 1.6-4$\cdot$10$^{-12}$\,erg\,cm$^{-3}$ for our regions.
To estimate the lower limit of energies available for magnetic reconnection and for simplicity of the calculations, we assume 
2$\cdot$10$^{-12}$\,erg\,cm$^{-3}$ as the total energy density of the turbulent magnetic field. We assume that about 90\% of this energy 
is lost during the expansion
of the magnetic field into the halo \citep{hanasz98}. This leaves 2$\cdot$10$^{-13}$\,erg\,cm$^{-3}$ that would be converted to thermal energy 
via the magnetic reconnection effects. Assuming an efficiency of 0.5 for the conversion of the magnetic energy to thermal energy, 
the energy of the hot gas with a density of around 10$^{-3}$\,cm$^{-3}$ (regions I and V, Table~\ref{totalhot}) 
would increase by $\sim$10$^{-10}$\,erg per particle, that is by 25-30\% (Table~\ref{particles}). 
If we consider region I, a decrease in the energy density of the magnetic field 
by 2$\cdot$10$^{-13}$\,erg\,cm$^{-3}$ would require  an initial strength of the magnetic field of around 8.5\,$\mu$G for this region. Although the 
derived values are still within the uncertainties, we note that these are lower limits, as we assumed above. 

Furthermore, these calculations 
could also support the idea of magnetic reconnection in region V, in which the vertical expansion of the magnetic fields should be easier.
The potential increase in the degree of order of the magnetic field would not be observed (see Sect.~\ref{hotfields}), but, similarly to region I, 
around 2$\cdot$10$^{-13}$\,erg\,cm$^{-3}$ of the magnetic field energy would be converted to additional $\sim$10$^{-10}$\,erg per particle of the hot gas.
Then, for both regions (I and V), reconnection heating would be responsible for the shift visible in Figs.~\ref{epmag} and~\ref{epbord}. 
The corresponding change in the temperature of the hot gas (by 25-30\%) would also be consistent with values 
derived for the other regions (Table~\ref{totalhot}). 

\section{Summary and Conclusions}
\label{cons}

In this paper we present an analysis of the radio and X-ray data for the spiral galaxy M\,101. Studies of the properties of the hot gas 
and the magnetic fields in different parts of the disk and the halo of the galaxy are performed.
Our findings can be summarised as follows:

\begin{enumerate}
\item For the first time, a high-resolution map of M\,101 at a high radio frequency was obtained. Because of the very large angular size of this galaxy,
        a single-dish map was needed to restore the large-scale emission lost in the interferometric observations.
\item   Although the majority of the polarised emission is associated with the galactic spiral arms, as suggested by the low-resolution observations 
        by \citet{berkhuijsen16}, region I, that is the inter-arm region between the massive eastern spiral arm and the galactic centre, 
        hosts a rudimentary magnetic arm.
\item The X-ray data show diffuse emission from the hot gas that extends beyond the eastern part of the galactic disk and suggests 
        interactions within the group environment. This seems to be confirmed by strong and moderately ordered magnetic fields in the massive 
        spiral arm, which hosts two prominent star-forming regions that could also be a product of enhanced star formation induced by these interactions.
\item The spectral analysis of the X-ray data suggests that most of the emission comes from the hot gas in the halo of M\,101. 
\item Three galactic regions, I, V, and S2$_{\rm X}$, show an increase in the temperature of the hot gas (and thus higher energies per particle), when 
        compared to other regions. However,  only in region I are both lower energy density and a higher order of the magnetic field found, 
	which is characteristic of magnetic arms. 
        We propose that these features can be explained with magnetic reconnection effects, which convert energy of the magnetic field
        to thermal energy of the hot gas.
\item   In region V, an area devoid of H$\alpha$ and \ion{H}{i} emission between the eastern spiral arm and the galactic centre, 
        our low-sensitivity spectrum suggests the existence of gas hotter than in the surrounding regions. This 
        could be a result of heating by magnetic reconnection. In this region, due to 
        a lack of any significant polarised radio emission that would suggest ordered magnetic fields, magnetic reconnection would need 
        to occur in the vertical component of the magnetic fields. As we argue in this paper, due to lower gas densities and the time required
        to start the magnetic reconnection, galactic halos should host the most efficient 
        reconnection heating, and which is thus easier to detect.
\end{enumerate}

Our results, together with our previous study of NGC\,6946 \citep{wezgowiec16} and M\,83 \citep{wezgowiec20}, suggest that the gas heating 
by magnetic reconnection might be possible to detect. More analyses, also of edge-on galaxies, 
are planned in order to give these detections a more statistical significance. 

\begin{acknowledgements}
We thank the anonymous referee for comments that helped to improve this paper. 
M.W., M.S., and M.U. are supported by the National Science Centre, Poland, with the grant project 2017/27/B/ST9/01050. 
M. H. acknowledges the support by the National Science Centre through the OPUS grant No. 2015/19/B/ST9/02959.
Research in the field at Ruhr University Bochum is supported by Deutsche Forschungsgemeinschaft SFB 1491.
The National Radio Astronomy Observatory is a facility of the National Science Foundation operated under cooperative agreement 
by Associated Universities, Inc. 
\end{acknowledgements}

\bibliographystyle{aa} 
\bibliography{myreferences} 

\end{document}